\pdfoutput=1

\documentclass[twocolumn,showpacs,aps,prd,superscriptaddress]{revtex4}

\usepackage[title]{appendix}
\usepackage{hyperref} 
\usepackage{graphicx}
\usepackage{dcolumn}
\usepackage{amsmath}
\usepackage{amssymb}
\usepackage{epsfig}
\usepackage{hhline}
\usepackage{soul}
\usepackage{tabularx,pifont,multirow}
\usepackage{subfigure}
\usepackage{colordvi}
\usepackage[usenames,dvipsnames]{xcolor}
\definecolor{taplum}{rgb}{0.67843, 0.49804, 0.65882}
\RequirePackage{xspace}
\usepackage{grffile}

\long\def\inst#1{\par\nobreak\kern 4pt\nobreak
  {\it #1}\par\vskip 10pt plus 3pt minus 3pt}

\def\lhcb {\mbox{LHCb}\xspace}
\def\lhc    {\mbox{LHC}\xspace}
\def\pt         {\mbox{$p_{\mathrm{ T}}$}\xspace}

\newcommand{\br}{\ensuremath{\mathcal{B}}\xspace}

\def\murad{\ensuremath{\mathrm{ \,\mu rad}}\xspace}
\def\mrad{\ensuremath{\mathrm{ \,mrad}}\xspace}
\def\mum  {\ensuremath{{\,\mu\mathrm{m}}}\xspace}
\def\mm   {\ensuremath{\mathrm{ \,mm}}\xspace}
\def\cm   {\ensuremath{\mathrm{ \,cm}}\xspace}
\def\m    {\ensuremath{\mathrm{ \,m}}\xspace}
\def\degk {\ensuremath {\mathrm{ \,K}}\xspace}
\newcommand{\kev}{\ensuremath{\mathrm{\,ke\kern -0.1em V}}\xspace}
\newcommand{\gev}{\ensuremath{\mathrm{\,Ge\kern -0.1em V}}\xspace}
\newcommand{\gevc}{\ensuremath{{\mathrm{\,Ge\kern -0.1em V\!/}c}}\xspace}
\newcommand{\mevcc}{\ensuremath{{\mathrm{\,Me\kern -0.1em V\!/}c^2}}\xspace}
\newcommand{\gevcc}{\ensuremath{{\mathrm{\,Ge\kern -0.1em V\!/}c^2}}\xspace}
\newcommand{\tev}{\ensuremath{\mathrm{\,Te\kern -0.1em V}}\xspace}

\def\ppsec{\ensuremath{\proton/\mathrm{s}}\xspace}
\def\mub{\ensuremath{{\mathrm{ \,\mu b}}}\xspace}
\def\geant      {\mbox{\textsc{Geant4}}\xspace}
\def\pythia     {\mbox{\textsc{Pythia}}\xspace}

\def\pr          {{\ensuremath{p}}\xspace}

\def\bxi{\ensuremath{\boldsymbol \xi}\xspace}
\def\bp{\ensuremath{\boldsymbol p}\xspace}

\def\Sb              {{\ensuremath{\rm S_2}}\xspace}

\def\W               {{\ensuremath{\rm W}}\xspace}

\newcommand{\pot}{\ensuremath{\mathrm{\,PoT}}\xspace}

\newcommand{\tevc}{\ensuremath{{\mathrm{\,Te\kern -0.1em V\!/}c}}\xspace}
\newcommand{\tevtevcccc}{\ensuremath{{\mathrm{\,Te\kern -0.1em V^2\!/}c^4}}\xspace}
\newcommand{\gevgevcc}{\ensuremath{{\mathrm{\,Ge\kern -0.1em V^2\!/}c^2}}\xspace}
\newcommand{\tevtevcc}{\ensuremath{{\mathrm{\,Te\kern -0.1em V^2\!/}c^2}}\xspace}

\def\Lc      {{\ensuremath{\Lz^+_\cquark}}\xspace}
\def\cquark    {{\ensuremath{\Pc}}\xspace}
\def\cquarkbar {{\ensuremath{\overline \cquark}}\xspace}
\def\ccbar     {{\ensuremath{\cquark\cquarkbar}}\xspace}

\def\Xicp    {{\ensuremath{\Xires^+_\cquark}}\xspace}
\def\Lz          {{\ensuremath{\PLambda}}\xspace}
\def\PLambda     {\ensuremath{\Lambda}\xspace}                 
\def\Pc      {\ensuremath{c}\xspace}
\def\Km      {{\ensuremath{\kaon^-}}\xspace}
\def\Kp      {{\ensuremath{\kaon^+}}\xspace}
\def\pip    {{\ensuremath{\pion^+}}\xspace}
\def\pim    {{\ensuremath{\pion^-}}\xspace}
\def\proton      {{\ensuremath{\Pp}}\xspace}
 \def\Pp      {\ensuremath{p}\xspace}
 \def\kaon    {{\ensuremath{\PK}}\xspace}
  \def\PK      {\ensuremath{K}\xspace}
  \def\pion   {{\ensuremath{\Ppi}}\xspace}
   \def\Ppi         {\ensuremath{\pi}\xspace}
   \def\piz    {{\ensuremath{\pion^0}}\xspace}
    \def\Ptau        {\ensuremath{\tau}\xspace}

\def\Sigmap    {{\ensuremath{\Sigma^+}}\xspace}
\def\Sigmam    {{\ensuremath{\Sigma^-}}\xspace}
\def\Xicp    {{\ensuremath{\Xi^+_\cquark}}\xspace}
\def\Xicz    {{\ensuremath{\Xi^0_\cquark}}\xspace}
\def\Omegacz    {{\ensuremath{\Omega^0_\cquark}}\xspace}
\def\Xim{\ensuremath{\Xi^-}\xspace}

\def\LcpKpi{\ensuremath{\Lc\rightarrow\proton\Km\pip}\xspace}

\newcommand{\eg}{\mbox{\itshape e.g.}\xspace}
\newcommand{\ie}{\mbox{\itshape i.e.}\xspace}

\begin{document}


\begin{flushleft}
\end{flushleft}


\title{
{
  \large \bf \boldmath
Progress towards the first measurement of charm baryon dipole moments
}
}

\author{S.~Aiola}
\affiliation{INFN Sezione di Milano, Milan, Italy} 
\author{L.~Bandiera}
\affiliation{INFN Sezione di Ferrara, Ferrara, Italy} 
\author{G.~Cavoto}
\affiliation{INFN Sezione di Roma, Rome, Italy} 
\affiliation{Universit\`a di Roma ``La Sapienza'', Rome, Italy}
\author{F.~De Benedetti}
\affiliation{INFN Sezione di Milano, Milan, Italy}
\author{J.~Fu}
\affiliation{INFN Sezione di Milano, Milan, Italy} 
\affiliation{Universit\`a degli Studi di Milano, Milan, Italy}
\author{V.~Guidi}
\affiliation{INFN Sezione di Ferrara, Ferrara, Italy} 
\affiliation{Universit\`a degli Studi di Ferrara, Ferrara, Italy}
\author{L.~Henry}
\affiliation{INFN Sezione di Milano, Milan, Italy} 
\affiliation{Universit\`a degli Studi di Milano, Milan, Italy}
\affiliation{IFIC, Universitat de Val\`encia-CSIC, Valencia, Spain}
\author{D.~Marangotto}
\affiliation{INFN Sezione di Milano, Milan, Italy} 
\affiliation{Universit\`a degli Studi di Milano, Milan, Italy}
\author{F.~Martinez Vidal}
\affiliation{IFIC, Universitat de Val\`encia-CSIC, Valencia, Spain}
\author{V.~Mascagna}
\affiliation{INFN Sezione di Milano Bicocca, Milan, Italy} \affiliation{Universit\`a degli Studi dell'Insubria, Como, Italy}
\author{J.~Mazorra de Cos}
\affiliation{IFIC, Universitat de Val\`encia-CSIC, Valencia, Spain}
\author{A.~Mazzolari}
\affiliation{INFN Sezione di Ferrara, Ferrara, Italy} 
\author{A.~Merli}
\affiliation{INFN Sezione di Milano, Milan, Italy} 
\affiliation{Universit\`a degli Studi di Milano, Milan, Italy}
\author{N.~Neri}
\affiliation{INFN Sezione di Milano, Milan, Italy} 
\affiliation{Universit\`a degli Studi di Milano, Milan, Italy}
\author{M.~Prest}
\affiliation{INFN Sezione di Milano Bicocca, Milan, Italy}
\affiliation{Universit\`a degli Studi dell'Insubria, Como, Italy}
\author{M.~Romagnoni}
\affiliation{INFN Sezione di Ferrara, Ferrara, Italy} 
\affiliation{Universit\`a degli Studi di Milano, Milan, Italy}
\author{J.~Ruiz Vidal}
\affiliation{IFIC, Universitat de Val\`encia-CSIC, Valencia, Spain}
\author{M.~Soldani\thanks{now at INFN Sezione di Ferrara, Ferrara, Italy}}
\affiliation{INFN Sezione di Milano Bicocca, Milan, Italy} \affiliation{Universit\`a degli Studi dell'Insubria, Como, Italy}
\author{A.~Sytov}
\affiliation{INFN Sezione di Ferrara, Ferrara, Italy} 
\author{V.~Tikhomirov}
\affiliation{Institute for Nuclear Problems and Belarusian State University, Minsk, Belarus} 
\author{E.~Vallazza}
\affiliation{INFN Sezione di Milano Bicocca, Milan, Italy}

\date{\today}

\begin{abstract}
Electromagnetic dipole moments of short-lived particles
are sensitive to physics within and beyond the Standard Model of
particle physics but have not been accessible experimentally to date. To perform such measurements it has been proposed to exploit the spin precession of channeled particles in bent crystals at the \lhc. 
Progress that enables the first measurement of charm baryon dipole moments is reported.
In particular, the design and characterization on beam of silicon and germanium
bent crystal prototypes, the optimization of the experimental setup, and advanced analysis techniques are discussed.
Sensitivity studies show that first measurements of \Lc and \Xicp baryon dipole moments can be performed 
in two years of data taking with 
an
experimental setup positioned upstream of the \lhcb detector.
\end{abstract}
\maketitle

%
\section{Introduction}
Electromagnetic dipole moments are static properties
of particles that are sensitive to physics within and beyond
the Standard Model (SM) of particle physics. 
For particles like the proton, neutron, muon and electron, such measurements
provide among the most stringent tests of the SM~\cite{Schneider:2017lff,Sahoo:2016zvr,Afach:2015sja,Bennett:2006fi,Bennett:2008dy,Hanneke:2008tm,Baron:2013eja}.
In classical physics, the magnetic dipole moment (MDM) measures the strength and
orientation 
of the magnetic field generated by the motion of electric charges or
by the intrinsic magnetism of matter.
In particle physics, the MDM is proportional to the particle
spin-polarization vector ${\boldsymbol s}$ and for spin-1/2 particles
is given by
${\boldsymbol\mu} = g\mu_\textrm{B}{\boldsymbol s}/2$ (Gaussian units), where
$g$ is the dimensionless gyromagnetic factor, $\mu_\textrm{B}=e\hslash/(2mc)$ is the particle magneton, and $m$ its mass. The spin polarization is a unit vector defined as
${\boldsymbol s}=2\langle{\boldsymbol S}\rangle/\hslash$, where ${\boldsymbol S}$
is the spin operator.
The measurement of the baryon MDM provides experimental anchor points for
low energy models of strong interactions,
while for leptons can be confronted with precise calculations
for a stringent SM test~\cite{Dainese:2019xrz,Miller:2007kk}.
The electric dipole moment (EDM) of a system
measures the separation of the positive and negative electric charge
distribution.
The particle EDM is proportional to its spin-polarization vector
and
is defined as ${\boldsymbol\delta} = d\mu_\textrm{B}{\boldsymbol s}/2$,
where $d$ is the gyroelectric factor. The EDM is expected to
be minuscule for baryons and leptons in the SM and any observation would
imply the existence
of physics beyond the SM~\cite{Chupp:2017rkp,Beacham:2019nyx,Unal:2020ezc}. 
There are no direct measurements to date of such properties for
charm and beauty baryons, and also for the $\tau$ lepton, due to the difficulties
imposed by their short lifetimes. For the charm quark (chromo-)EDM only indirect limits exist based on the experimental bounds on the neutron EDM~\cite{Sala:2013osa,Gisbert:2019ftm}.

Recently it has been proposed to measure charm baryon MDM/EDM at a 
fixed-target experiment to be installed at the Large Hadron Collider
(LHC)~\cite{Baryshevsky:2016cul,Burmistrov:2194564,Botella:2016ksl,Bagli:2017foe,Bezshyyko:2017var}, following an experiment on the \Sigmap hyperon MDM~\cite{Chen:1992wx,SAMSONOV1996271} and other proposals at Fermilab~\cite{Baublis:1994ku,Khanzadeev:1996ag}.
By exploiting the phenomenon of particle channeling in bent crystals,
the electric and magnetic dipole moments of short-lived particles
can be measured by studying the spin precession induced by the intense
electric field between the crystal atomic planes, firstly proposed
by Baryshevsky in 1979~\cite{Baryshevsky:1979}.
In a crystal, the strong electric field experienced by a positively charged particle
in the proximity of the atomic planes exerts a strong force
and the particle trajectory becomes confined within two crystalline planes.
This phenomenon, called planar channeling, can occur if the entrance angle between
the particle trajectory and a crystal plane is lower than the Lindhard
critical angle for channeling, 
$\theta_L =\sqrt{ 2 U_0/(p\beta c)}$, where $U_0$ is the potential-well depth,
$p$ the particle momentum and $\beta$ its velocity~\cite{Lindhard,Biryukov1997}.
For a 1\tev charged particle $\theta_L \approx 6.3 \murad$ ($\theta_L \approx 7.7 \murad$) in a silicon (germanium) crystal.
%
%
\begin{figure}[h]
\begin{center}
\includegraphics[width=0.90\columnwidth]{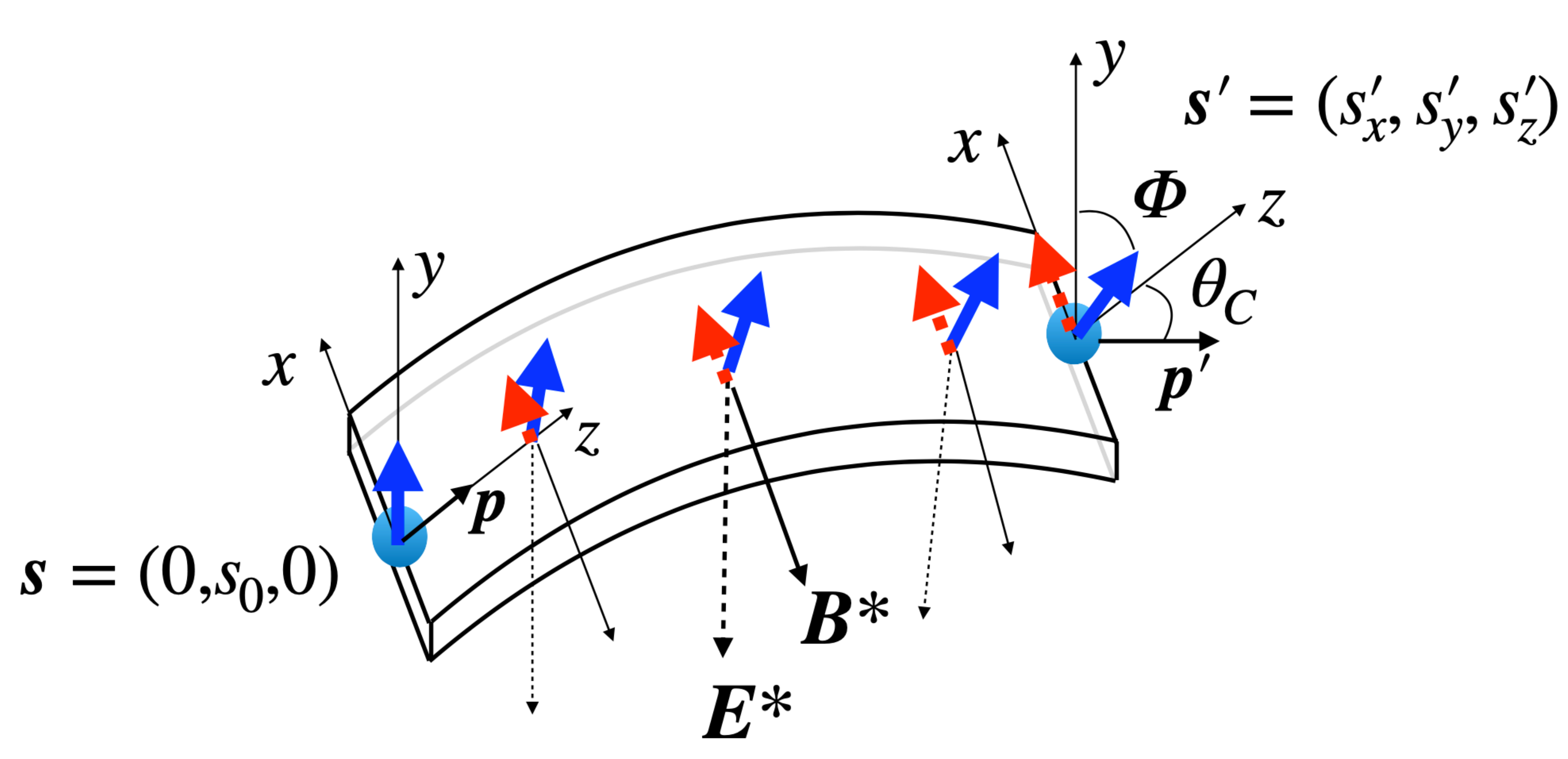}
\caption{\label{fig:SpinPrecession_region}
Deflection of a particle with initial and final momentum \bp and $\bp'$,
and spin-polarization precession in the $yz$ and $xz$ plane induced by the
MDM and the EDM, respectively.
The solid (blue) arrow represents the initial spin-polarization vector ${\boldsymbol s}$, aligned along the $y$ axis in this example, normal to the crystal plane at entry. The spin is rotated of an angle $\varPhi$ to ${\boldsymbol s'}$ after channeling in a crystal with bending angle
$\theta_C$.
The dashed (red) arrow indicates the (magnified) $s'_x$ spin
polarization component proportional to the particle EDM.
The ${\boldsymbol B^*}$ (${\boldsymbol E^*}$) indicates
the magnetic (electric) field in the particle rest frame.
}
\end{center}
\end{figure}
The \Lc and \Xicp charm baryons, produced by interactions of the
7\tev LHC proton beam on a fixed target,  
are allowed to have
initial polarization
perpendicular to the production plane,
due to parity symmetry conservation in strong interactions.
The generation occurs within a cone with aperture $1/\gamma\approx 1\mrad$,
where $\gamma$ is the Lorentz factor of the charm baryon.
After the target, a bent crystal is placed to act as a special type of  spectrometer
deflecting channeled particles of a constant angle $\theta_C$ within the detector acceptance. The upgraded \lhcb detector
is particularly suited for this experiment thanks to its forward geometry and excellent
performance for the reconstruction of heavy hadrons~\cite{Aaij:2014jba,Bediaga:2012uyd}.
However, only a small fraction of the produced charm baryons, entering the crystal at an
angle within the critical angle, is channeled and deflected. 
A sizeable spin precession, induced by the intense electromagnetic
field between crystal atomic planes, allows to probe for 
electromagnetic dipole moments, as illustrated in
Fig.~\ref{fig:SpinPrecession_region}.
For ultra-relativistic channeled particles with $\gamma \gg 1$,
the spin precession angle induced by the MDM in the $yz$ plane is
\begin{equation}
\label{eq:varphi}
\varPhi \approx \frac{g-2}{2}\gamma \theta_C ,
\end{equation}
where $\theta_C= L/\rho$ is the crystal bending angle, $L$ is the length of the
 crystal arc and $\rho$ the curvature radius.
 The presence of a non-zero EDM, in the limit of $d\ll g-2, $ introduces a spin-polarization component
 $s'_x$ perpendicular to the plane of the crystal bending~\cite{Botella:2016ksl},
\begin{equation}
s'_x \approx s_0 \frac{d}{g-2}(1-\cos \varPhi),
\end{equation}
where $s_0$ is the initial polarization.
 According to previous studies,
 a germanium crystal allows for enhanced sensitivity to MDM and EDM
with respect to the silicon crystal~\cite{Bagli:2017foe,Fomin:2019wuw}
 due to the higher electric field between crystal atomic planes.

In this article, the design and characterization on beam
of the first silicon and germaniun bent crystal prototypes for charm baryon
spin precession are reported in Sec.~\ref{sec:crystals}.
Advanced experimental techniques and sensitivities studies for optimal future measurements
of charm baryon dipole moments are discussed in Sec.~\ref{sec:expTech} and Sec.~\ref{sec:sensitivity}, respectively.
The main improvements with respect to previous studies~\cite{Bagli:2017foe}
include:
a full amplitude analysis for optimal determination of baryon polarization,
a more realistic polarization model~\cite{Bezshyyko:2017var},
an optimized target thickness,
CRYSTALRAD Monte Carlo simulations for particle channeling also for cryogenic temperatures~\cite{Sytov:2014jfa},
and additional charm baryon decay modes.
%

%
\hfill \break
\section{Bent crystal prototypes and test on beam}
\label{sec:crystals}
Crystal-assisted efficient steering of high-energy particle beams
requires bent crystals with a tight control over their
deformational state and an exceptionally low number of crystalline defects.
Non-uniformities of the deformational state of the crystal and crystalline
defects would indeed lead particles
to be lost from channeling condition, lowering the efficiency of the
steering process~\cite{Biryukov1997,Bagli:2015jga}.

Crystals are manufactured through a revisitation of a
protocol~\cite{Baricordi:2007zz,Germogli:2015gja,Mazzolari:2018hsu}
already assessed for manufacturing of crystals of few \mm length suitable for steering
of particle beams circulating at the LHC~\cite{Scandale:2016krl}.
The prime materials are a (111) oriented $5 \mm$ thick silicon wafer
 and a (110) oriented $1 \mm$ thick germanium wafer.
With the aim to maintain an optimal steering efficiency, wafers with less than
$1/\cm^2$ dislocation density were selected from a large stock of wafers.

\begin{figure*}[htb!]
\begin{center}
\includegraphics[width=0.7\columnwidth]{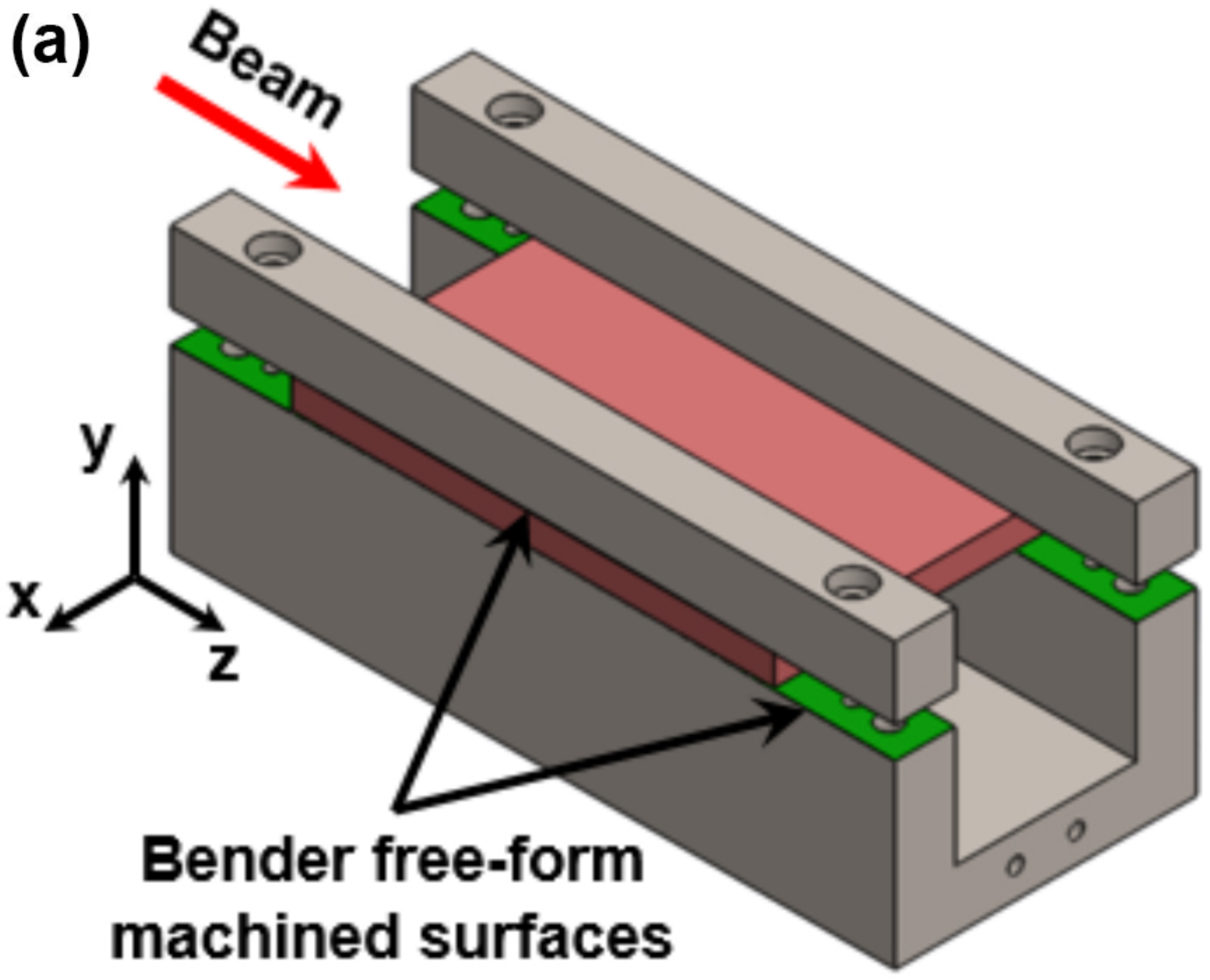}
\includegraphics[width=0.7\columnwidth]{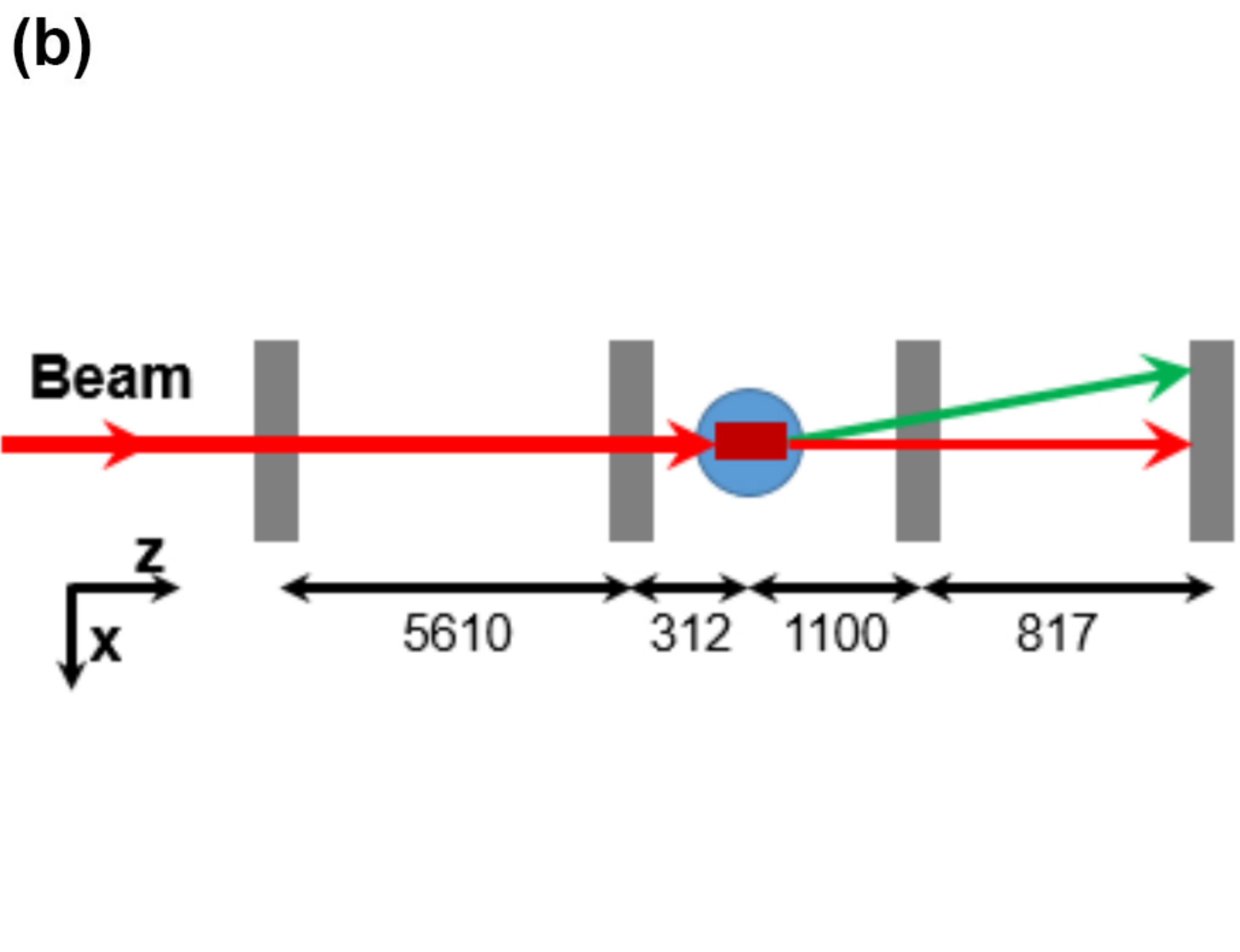}
\caption{\label{fig:crystal_sketch}(a) Sketch of an assembly made of crystal (red part) and associated bender (grey parts). Surfaces in contact with the crystal (green color) have been machined to a free-form surface. As the crystal is clamped in the bender, it bends around the $x$ axis. A particle beam,
initially propagating along the $z$ axis is channeled between bent atomic
planes and deflected. (b) Sketch of the experimental setup at the CERN H8
beam line. A $180\gev$ positive hadron beam is directed to the crystal.
The trajectory of each particle is reconstructed before and after interaction with the crystal by a tracking telescope (grey boxes). The crystal (red rectangle) is mounted on a high-resolution goniometer (blue circle). As the crystal is oriented in order to channel the incoming beam between its atomic planes, beam steering occurs and a fraction of the beam is deflected (green arrow). Distances between elements of the setup are expressed in \mm.}
\end{center}
\end{figure*}

Mechanical dicing was applied to  wafers to obtain a $50\times5\times80 \mm^3$
crystal and a $35\times1\times55 \mm^3$ crystal made of silicon and germanium,
respectively.
The first value indicates the dimension along the beam and in both cases 
was chosen according to Ref.~\cite{Bagli:2017foe}.
Each crystal was mechanically bent along
 this longer dimension through a bender properly shaped to impart to the
 crystal the nominal bending radius
of $5.0\m$ for silicon and $3.6\m$ for germanium, respectively.
Given the nominal bending radius,
channeled particles are deflected at an angle equal to $16\mrad$ for silicon
and $15\mrad$ for germanium, see Fig.~\ref{fig:crystal_sketch}.
The uniformity of the crystal deformation plays a key role for obtaining the
expected steering efficiency: to enhance
uniformity of crystal curvature, the nominal shape of the surfaces of the bender
in contact with the crystal have been optimized through finite element models
to free-form surfaces.
After bending, the crystal deformational state was characterized~\cite{GERMOGLI2017308} by means of high-resolution diffraction
of a $8\kev$ X-ray beam.

%
%

At the H8 external beam line of the Super Proton Synchrotron (SPS) at CERN,
the crystal/bender assemblies were mounted on a goniometer capable of
rotations with accuracy of $1 \murad$ and aligned to a 
 $180\gev$ positively charged hadron beam in order to measure particle channeling.
The beam angular divergence ($26\murad$) resulted to be wider than
$\theta_L$ for $180\gev$ particles ($\approx 15\murad$ for silicon and $\approx 17\murad$ for germanium), also the
 dimension along the $y$ coordinate ($\approx 8\mm$) 
 was larger than the geometrical size of the crystal.
 To select particle trajectories intercepting the crystal and
 impinging on bent atomic planes at angle comparable to $\theta_L$,
 a tracking telescope based on four microstrips detectors was used.
 Two detectors were placed before and two were placed after the crystal to
 reconstruct particle trajectories before and after the interaction.
  Monte Carlo simulations of the beam test setup, based on the \geant toolkit~\cite{Agostinelli2003250},
  are used to determine the angular uncertainty on particle trajectories
  before and after the interaction with the crystal,
  which resulted to be $\approx 7 \murad$ and $\approx 50 \murad$, respectively.
\begin{figure*}[htb!]
\begin{center}
{\begin{tabular}{cc}
\subfigcapskip = 5pt
\subfigure[]{\includegraphics[width=0.45\textwidth]{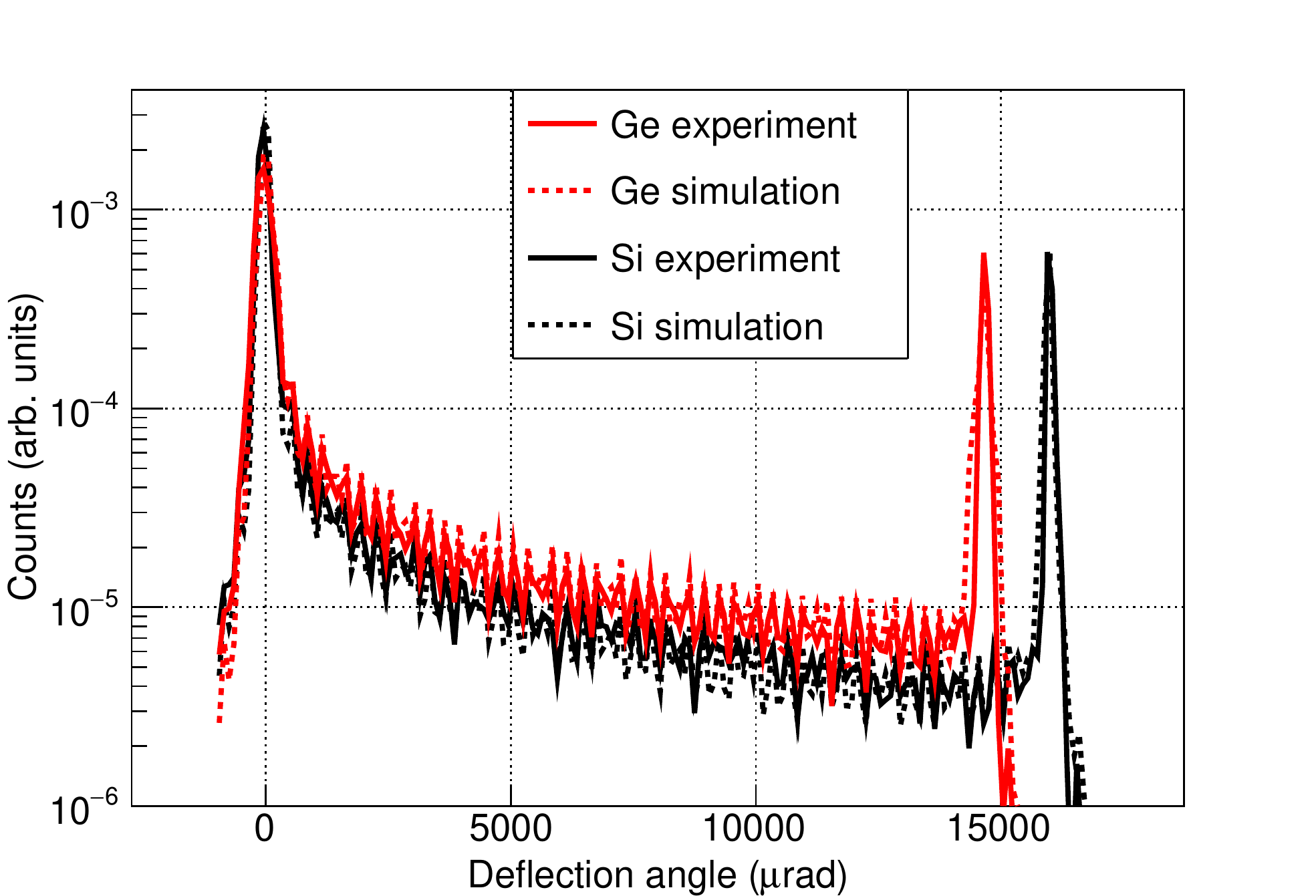}}
\subfigure[]{\includegraphics[width=0.45\textwidth]{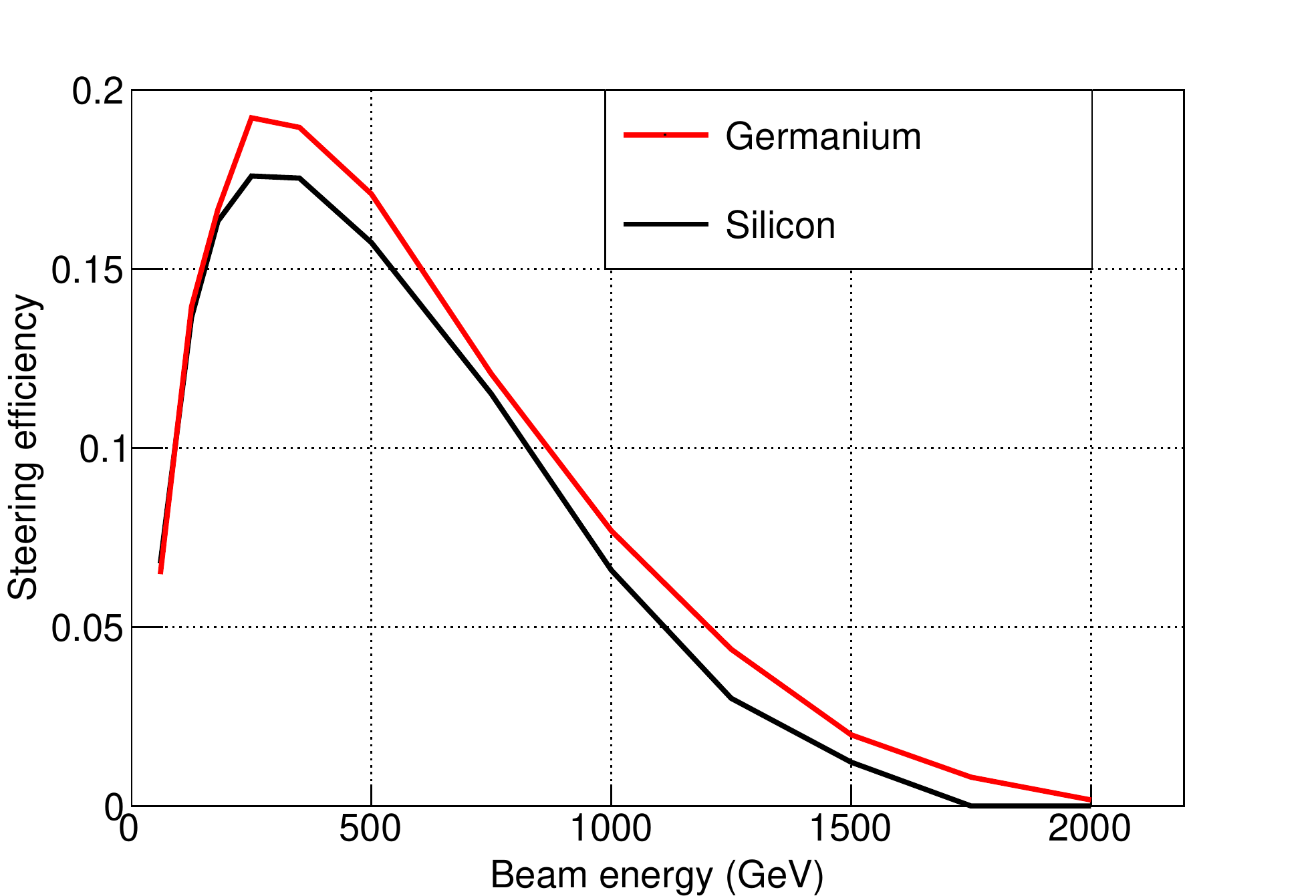}}
\end{tabular}}
\caption{\label{fig:deflection_angle}
(a) Angular distributions of a $180\gev$ hadron beam after interaction with the silicon (black lines) and germanium (red lines) crystals.
Experimental data (solid curves)
are compared to results of Monte Carlo simulations (dashed curves).
The maxima located on the right side of the plot correspond to particles that
are deflected, being channeled over the entire crystal length. The oscillations between the two peaks  are due to the discrete structure of the strip detectors. By selecting particles impinging on the crystal within $\pm30 \murad$,
 steering efficiencies of $(8.9\pm0.5)\%$  and $(10.8\pm0.5)\%$ are measured
 for silicon and germanium crystals, respectively. (b) Monte Carlo simulation results for the steering
 efficiency of the two crystals as a function of the energy of the beam, assuming uniform bending radius and a particle beam with uniform angular distribution at the crystal entrance.
}
\end{center}
\end{figure*}

The angular distribution of the beam after interaction with the crystal
is shown in Fig.~\ref{fig:deflection_angle}(a).
 The peak of the distribution on the left is populated by particles which
 are not channeled between atomic planes at the crystal entry
 face~\cite{TARATIN1987425}
 and by particles which are initially channeled but with an impact parameter
 with respect to atomic planes smaller than the thermal vibration amplitude
 of the atoms of the crystal. For such particles, single scattering events
 with the inner shell electrons of the atoms or with the atomic nuclei
 result in a drastic change of their trajectory, leading them out of
 channeling regime soon after being channeled~\cite{Scandale:2015zea}.
 The peaks on the right, populated by channeled particles over the entire
 length of the crystal, are centered
 at angles of $15988\pm5 \murad$ and
 $14670\pm2 \murad$ for the case of silicon and germanium crystals,
 respectively.
 Given the fact that in the final application the crystal would interact with a particle beam whose divergence is much larger than the critical angle for channeling, the steering efficiency of the crystal is measured for particles reaching the crystal at a nominal angle of $\pm 30 \murad$, corresponding to an angular window about two times larger than the critical angle for channeling, and with deflection  angle in a window of $\pm 500 \murad$ around the peak of maximum deflection. Within that angular window, the efficiencies are measured according to Ref.~\cite{ROSSI2015369}.
\begin{table*}
  \begin{tabular}{lcccccc}
    \hline
\multirow{2}{*}{} &
      \multirow{2}{*}{Length (\mm)} & 
      \multirow{2}{*}{Thickness (\mm)} &   
      \multicolumn{2}{c}{\rule{0pt}{2.4ex}Deflection angle (\mrad)} &
      \multicolumn{2}{c}{Steering efficiency (\%)}\\
     &  &  & X-Ray & Channeling & Measured & Simulation \\
    \hline
    \rule{0pt}{2.4ex}Germanium & 55  & 1 & $14.5 \pm 0.8$ & $14.670\pm0.002$ & $10.8\pm0.5$ & $12.3\pm0.5$ \\
                       Silicon & 80 & 5 & $16.1\pm 0.8$ & $15.988\pm0.005$ & $8.9\pm 0.5$ & $9.9\pm 0.5$ \\
    \hline
  \end{tabular}
  \caption{Geometrical parameters of silicon and germanium crystals used in the experiment, and recorded channeling efficiencies compared to expected from Monte Carlo simulations. The length (thickness) represents the longitudinal (transverse) dimension of the crystal with respect to the beam direction. The uncertainties on the steering efficiency results take into account both statistical and systematic effects.}
  \label{table:crystals}
\end{table*}
Table~\ref{table:crystals} summarizes the most important features of the 
crystal prototypes and the measured channeling efficiencies. 
Using Monte Carlo simulations for the interaction between particle
 beam and crystal (for details see Appendix~\ref{sec:methods_simuinteractions}),
the steering efficiency as a function of the energy of the beam is determined
for the two crystals assuming a uniform bending radius.
Steering efficiency results are shown in Fig.~\ref{fig:deflection_angle}(b).
%

\hfill \break
\section{Experimental techniques}
\label{sec:expTech}

Charm baryon MDM and EDM measurements require advanced analysis techniques discussed below.
Combining these with an optimized experimental setup, the sensitivity to the observables of interest is extracted. To that end, large samples of pseudoexperiments are used, where the kinematics of \Lc and \Xicp baryons produced in the target are generated according to their momentum spectra
as obtained from \pythia 8.244 event
generator~\cite{Sjostrand:2014zea}
tuned to NNPDF3.1sx NNLO NLLx LUXQED $\alpha_s(m_Z)=0.118$
parton distribution functions~\cite{Ball:2017nwa,Bertone:2017bme} including \lhcb charm production data~\cite{Bertone:2018dse}.

\begin{figure}[h]
\begin{center}
\includegraphics[width=0.90\columnwidth]{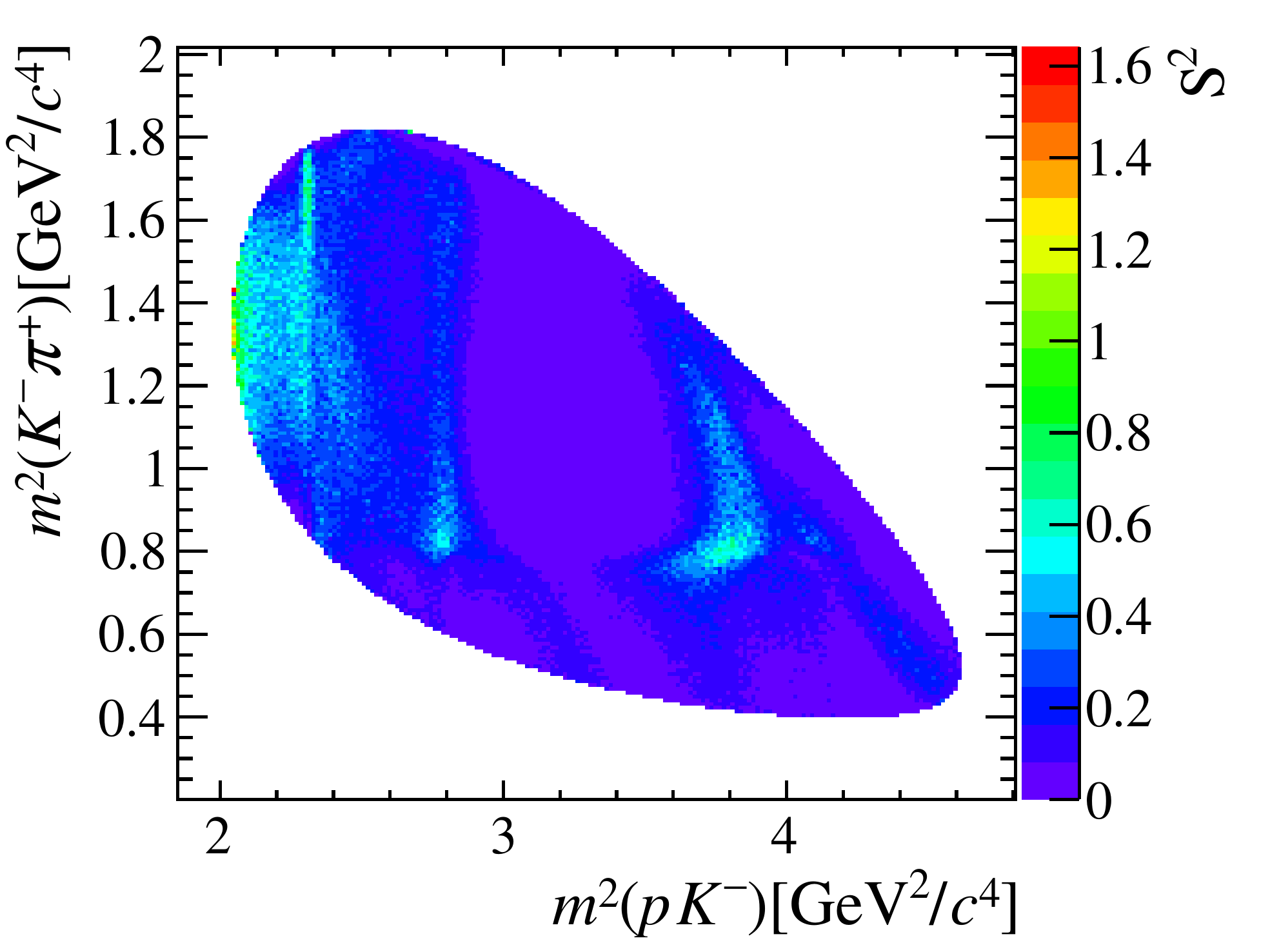}
\caption{\label{fig:P_sensitivity} Average event information $S^2$ to the \Lc baryon
polarization as a function of the Dalitz plot position
defined by the squared invariant masses $m^2(\proton\Km)$ and $m^2(\Km\pip)$, as obtained from
a large statistics pseudoexperiment.
}
\end{center}
\end{figure}

As a case study, the $\Lc\rightarrow\proton\Km\pip$ baryon decay is discussed.  
In the helicity formalism~\cite{Jacob:1959at}
the transition amplitude between an initial state with spin projection $m_\Lc$ along the quantization axis and a final state with proton helicity $m_\proton$ is indicated as ${\cal A}_{m_\Lc,m_\proton}(\bxi)$,
as a function of the phase space variables \bxi.
The decay distribution can be written  as
\begin{equation}
\label{eq:W}
W(\bxi|s) = f(\bxi)+s g(\bxi),
\end{equation}
where $s$ is the spin-polarization magnitude
 along a given axis, and  
the functions $f(\bxi)$ and $g(\bxi)$ are determined by the
decay amplitudes ${\cal A}_{m_\Lc,m_\proton}(\bxi)$ as
\begin{eqnarray}
f(\bxi) &=& W(\bxi|s=0) \nonumber \\
        &\propto& \sum_{m_\proton=\pm 1/2}\left[\left|{\cal A}_{1/2,m_\proton}(\bxi)\right|^2+ 
                   \left|{\cal A}_{-1/2,m_\proton}(\bxi)\right|^2\right], \nonumber
\end{eqnarray}
\begin{eqnarray}
\label{eq:fg}                  
g(\bxi) &=& \frac{1}{2} \big[ W(\bxi|s=1)-W(\bxi,s=-1) \big] \\
        &\propto& \sum_{m_\proton=\pm 1/2}\left[\left|{\cal A}_{1/2,m_\proton}(\bxi)\right|^2-
                   \left|{\cal A}_{-1/2,m_\proton}(\bxi)\right|^2\right], \nonumber 
\end{eqnarray}
and satisfy the normalization conditions
\begin{equation}
\label{eq:fg_int}
\int f(\bxi)d\bxi=1, \quad \quad \quad \int g(\bxi)d\bxi =0.
\end{equation}
The value of the average event information $S^2$ represents the sensitivity 
to the polarization $s$ that can be computed as~\cite{Davier:1992nw}
\begin{equation}
\label{eq:pol_sens}
S^2=\int \frac{g^2(\bxi)}{f(\bxi)+s_0 g(\bxi)}d\bxi,
\end{equation}
where $s_0$ is the best estimate for the polarization whose variance is $\sigma_{s}^2 = \left( N S^2 \right)^{-1}$, with $N$ the number of signal events.
The $\Lc\rightarrow\proton\Km\pip$ decay amplitude is  dominated by the sum of quasi two-body $pK^{*}$, $\Lambda^{*} \pip$, and $\Delta^{++}\Km$ resonant contributions~\cite{Tanabashi:2018oca,Aitala:1999uq}.
The average event information for $s_0=0$, based on a pseudoexperiment,  is shown in Fig.~\ref{fig:P_sensitivity} as a function of the Dalitz plot position,
defined by the squared invariant masses
$m^2(\proton\Km)$ and $m^2(\Km\pip)$,
with an average value $S^2\approx0.145$~\cite{Marangotto:2020tzf}.
This result can be compared with a two-body decay, where
$S^2$ for $s_0=0$ is related to the effective decay
asymmetry parameter $\alpha_{\rm eff}$, which characterizes parity violation,
$S^2 = \alpha_{\rm eff}^2 /3$,
and it is found to be compatible to that estimated for the
$\Lc\rightarrow \Delta^{++}\Km$ decay using the known value  
$\alpha_{\Delta^{++}\Km}=-0.67\pm0.30$~\cite{Botella:2016ksl}, affected by large
uncertainty.
Hence, the full amplitude analysis of the \LcpKpi decay enables 
the measurement of the \Lc spin precession using three-body decay modes~\cite{Marangotto:2020tzf,Marangotto:2020ead}, which
effectively increases the statistics for the dipole moments measurement
by a factor \mbox{${\cal B}(\Lc\rightarrow\proton\Km\pip)/{\cal B}(\Lc\rightarrow\Delta^{++}\Km)\approx 6$} with respect to that assumed in
previous studies~\cite{Botella:2016ksl,Bagli:2017foe,Bezshyyko:2017var,Fomin:2019wuw}.

Additional \Lc three- and four-body decays can be exploited for the measurement of the dipole moments
in order to further increase the signal yield up to an additional factor of three, assuming similar average event
information $S^2$ for all decays. This can be done analogously for \Xicp baryon decays.
In Tables~\ref{tab:Lc_decay_modes} and~\ref{tab:Xic_decay_modes} are listed
the decay modes considered for the sensitivity studies
with corresponding effective branching fractions,
taking into account the ability to reconstruct the final state charged particles with the \lhcb detector.
Due to the high momentum of charm baryons produced in the fixed-target interactions, a large fraction of the long-lived $\Sigmap$,
$\Sigmam$ and $\Xim$ strange baryons 
traverse the entire \lhcb tracking system before decaying,
thus their trajectories can be reconstructed
as for stable charged particles.
The crystal bending angle and length determine the momentum
spectrum of channneled charm baryons and consequently the fraction of strange
 baryons reconstructed as stable charged particles.
This requires that final state particles
travel more than $9.4\m$ from their production point
positioned $1.2\m$ upstream of the nominal \proton\proton interaction region.
Despite the lower production rate of the \Xicp compared to the \Lc baryon, estimated to be about 70\%~\cite{Bagli:2017foe},
and its smaller effective branching ratio, similar signal yields are expected for both baryon types decaying after the bent crystal
due to their different lifetimes, about a factor of two longer for the \Xicp baryon.

When there is a $\piz$ meson in the final state, a partial reconstruction of the decay can be performed based on charged particles only (for details see Appendix~\ref{sec:methods_reco}).
The charm baryon invariant mass
is reconstructed using the corrected mass~\cite{Abe:1997sb},
the charm hadron flight length and direction are determined by reconstructing
the production and the decay vertex positions, and its momentum can be
estimated without bias by reconstructing the decay kinematics.
Similarly, the average charm baryon spin-polarization vector can also
be measured with no bias employing the reconstructed kinematics,
and ultimately using a technique based on templates for full positive
and negative polarization, similar to that for the measurement of the polarization of the \Ptau leptons where the neutrinos
are not reconstructed~\cite{Davier:1992nw,Fu:2019utm}.
\begin{table}[htb]
{\begin{tabular}{lccc}
\hline
\rule{0pt}{2.4ex}\Lc final state                & ${\cal B}$ (\%)      & $\epsilon_{\textrm{3trk}}$ & ${\cal B}_{\textrm{eff}}$ (\%)      \\
\hline
 \rule{0pt}{2.4ex}$\proton\Km\pip$              &  $6.28\pm 0.32$      &  0.99    &  6.25  \\   
 $\Sigmap\pim\pip$                              &  $4.50\pm 0.25$      &  0.54    &  2.43    \\
 $\Sigmam\pip\pip$                              &  $1.87\pm 0.18$      &  0.71    &  1.33    \\
 $\proton\pim\pip$                              &  $0.461\pm 0.028$    &  1.00    &  0.46    \\ 
 $\Xim\Kp\pip$                                  &  $0.62\pm 0.06$      &  0.73    &  0.45    \\
 $\Sigmap\Km\Kp$                                &  $0.35\pm 0.04$      &  0.51    &  0.18    \\
 $\proton\Km\Kp$                                &  $0.106\pm 0.006$    &  0.98    &  0.11    \\
 $\Sigmap\pim\Kp$                               &  $0.21\pm 0.06$      &  0.54    &  0.11    \\
\hline
 \rule{0pt}{2.4ex}$\proton\Km\pip\piz$          &  $4.46\pm 0.30$      &  0.99    &  4.43    \\
 $\Sigmap\pim\pip\piz$                          &  $3.20$              &  0.54    &  1.72    \\
 $\Sigmam\pip\pip\piz$                          &  $2.1\pm0.4$         &  0.71    &  1.49    \\
 \hline
\rule{0pt}{2.4ex}$\Sigmap[\proton\piz]\pim\pip$ &  $2.32$              &  0.46    & 1.06    \\
$\Sigmap[\proton\piz]\Km\Kp$                    &  $0.18$              &  0.46    & 0.08    \\
$\Sigmap[\proton\piz]\pim\Kp$                   &  $0.11$              &  0.46    & 0.05    \\
\hline
\rule{0pt}{2.4ex}All   & - & -  &   20.2 \\
\hline
\end{tabular}}
\caption{List of \Lc baryon decay modes and corresponding absolute branching fraction ${\cal B}$~\cite{Tanabashi:2018oca},
efficiency for \lhcb reconstructibility of the three-charged particles $\epsilon_{\textrm{3trk}}$,
and the effective branching fraction ${\cal B}_{\textrm{eff}}={\cal B}\times\epsilon_{\textrm{3trk}}$.
The efficiency and effective branching fraction depend on the crystal parameters, here for germanium at room temperature with deflection angle of 16\mrad and 10\cm length.
}
\label{tab:Lc_decay_modes}
\end{table}
\begin{table}[htb]
{\begin{tabular}{lcccc}
\hline
\rule{0pt}{2.4ex}\Xicp final state               &  ${\cal RB}$      & ${\cal B}$ (\%)  & $\epsilon_{\textrm{3trk}}$ & ${\cal B}_{\textrm{eff}}$ (\%)      \\
\hline
\rule{0pt}{2.4ex}$\Xim\pip\pip$                  & 1                 & $2.86\pm1.27$    &  0.64  &  1.84    \\
$\Sigmap\Km\pip$                                 & $0.94\pm 0.10$    & -                &  0.42  &  1.14    \\
$\Sigmap\pim\pip$                                & $0.48\pm 0.20$    & -                &  0.44  &  0.60   \\
$\proton\Km\pip$                                 & $0.21\pm 0.04$    & -                &  0.99  &  0.60   \\
$\Sigmam\pip\pip$                                & $0.18\pm 0.09$    &  -               &  0.61  &  0.31    \\
$\Sigmap\Km\Kp$                                  & $0.15\pm 0.06$    &  -               &  0.41  &  0.18    \\
$\Omega^-\Kp\pip$                                & $0.07\pm 0.04$    &  -               &  0.42  &  0.08    \\
 \hline
\rule{0pt}{2.4ex}$\Sigma^+[\proton\piz]\Km\pip$  &  0.48             &  -               &  0.57  &  0.79    \\
$\Sigma^+[\proton\piz]\pim\pip$                  &  0.25             &  -               &  0.57  &  0.40    \\
$\Sigma^+[\proton\piz]\Km\Kp$                    &  0.08             &  -               &  0.59  &  0.13    \\
 \hline
\rule{0pt}{2.4ex}All   & -  & -  & -  &   6.1 \\
\hline
\end{tabular}}
 \caption{List of \Xicp baryon decay modes and corresponding relative branching 
absolute \mbox{$\Xicp\to\Xim\pip\pim$} branching fraction ${\cal B}$~\cite{Li:2019atu},
estimate of the efficiency for \lhcb reconstructibility of the three-charged particles $\epsilon_{\textrm{3trk}}$,
and the effective branching fraction ${\cal B}_{\textrm{eff}}={\cal RB}\times{\cal B}(\Xicp\to\Xim\pip\pim)\times\epsilon_{\textrm{3trk}}$ or
${\cal B}_{\textrm{eff}}={\cal B}\times\epsilon_{\textrm{3trk}}$ where ${\cal B}$ measurement is available.
The reported $\epsilon_{\textrm{3trk}}$ and ${\cal B}_{\textrm{eff}}$ are for a 16\mrad bent, 10\cm long germanium crystal at room temperature.
}
\label{tab:Xic_decay_modes}
\end{table}

The charm baryon polarization is perpendicular to the production plane, defined
by the momenta of the impinging proton and that of the outgoing \Lc baryon,
and is parameterized as a function of its transverse momentum \pt with respect to the direction of the impinging proton as
\begin{equation}
\label{eq:pol_pt}
s_0(\pt)\approx A\left(1-e^{-B p^2_T}\right),
\end{equation}
with $A \approx -0.9$ and $B \approx 0.4$~$(\gevc)^{-2}$.
The parametrization is based on a phenomenological dependence~\cite{Barlag:1994bk}
used to describe the experimental results~\cite{Aitala:1999uq}.

The \Lc and \Xicp baryon polarization versus \pt can be measured precisely in
fixed-target collisions at \lhcb using the SMOG system~\cite{Aaij:2014ida,Bursche:2649878} to further improve the polarization model.

\hfill \break
\section{Sensitivity studies and setup optimization}
\label{sec:sensitivity}

The sensitivity to dipole moments is studied using pseudoexperiments and \pythia simulations for different angles between the impinging proton direction and the crystal orientation.
The crystal reference frame $(x, y, z)$ is rotated at different angles $\theta_{y,C}$ around the $x_L$ axis in the $(x_L, y_L, z_L)$ laboratory frame~\cite{Fomin:2019wuw},
with $x$ parallel to the $x_L$ axis and $z_L$ parallel to the impinging proton direction.  
For non-zero $\theta_{y,C}$ values an initial polarization along the  $x$ and $y$ axes is induced, 
\begin{equation}
\label{eq:pol_init}
{\boldsymbol s} = (s_x,s_y,0) \approx \frac{s_0(\pt)}{\pt} \left( -p_{y_L}, p_{x_L}, 0 \right),
\end{equation}
where $p_{x_L}$ and $p_{y_L}=p\sin\theta_{y,C}$ are the transverse momentum components along the laboratory $x_L$- and $y_L$-axis, respectively, and $p$ 
is the total momentum of the charm baryon.
This is a consequence of parity conservation in strong interactions that forces the \Lc polarization vector to be perpendicular to the production plane, defined by the proton and the 
\Lc baryon momenta.
A probability density function based upon the quasi-two body approximation for the decay rate,
\begin{equation}
{\cal W} \propto 1 + \alpha_{\rm eff} {\boldsymbol s}' \cdot \hat{\boldsymbol k},
\end{equation}
is used in order to effectively reproduce the fit procedure for data,
where $\hat{\boldsymbol k}$ is the 
direction of the resonant intermediate state in the charm baryon helicity frame
and ${\boldsymbol s}'$ the spin-polarization vector after spin precession in the crystal,
\begin{align}
  s'_x & \approx   s_y \frac{a'd'}{{a'}^2_d} (1-\cos\varPhi) +
  s_x\left(\frac{{a'}^2}{{a'}_d^2}+\frac{{d'}^2}{{a'}^2_d}\cos\varPhi\right) ,
  \nonumber \\
  s'_y & \approx   s_y \left(\frac{{d'}^2}{{a'}^2_d}+\frac{{a'}^2}{{a'}_d^2}\cos\varPhi\right)                   + s_x \frac{a'd'}{{a'}^2_d} (1-\cos\varPhi) ,
  \nonumber \\
  s'_z & \approx  -s_y \frac{a'}{{a'}_d}\sin\varPhi
  + s_x \frac{d'}{{a'}_d} \sin\varPhi ,
\label{eq:eom_sxsy_full}
\end{align}
where $a'=a+1/(1+\gamma)$, with $a=(g-2)/2$ the anomalous magnetic moment, $d'=d/2$, $a'_d=\sqrt{{a'}^2+{d'}^2}$,
and $\varPhi = \gamma \theta_C a'_d$.
These expressions hold at ${\cal O}(10^{-2})$ precision and follow by solving the 
T-BMT equation~\cite{Thomas:1926dy,Thomas:1927yu,Bargmann:1959gz} 
for initial polarization along the $x$ and $y$ crystal frame directions, according to the procedure described elsewhere~\cite{Botella:2016ksl,Bagli:2017foe}.
The expected precession angle $\varPhi$ 
and the initial polarization-vector ${\boldsymbol s}$ can
be estimated on event-by-event basis using
the measured boost factor $\gamma$ and the transverse momentum of the charm baryon.

\begin{figure}[htb!]
\begin{center}
\subfigure[]{\includegraphics[width=0.75\columnwidth]{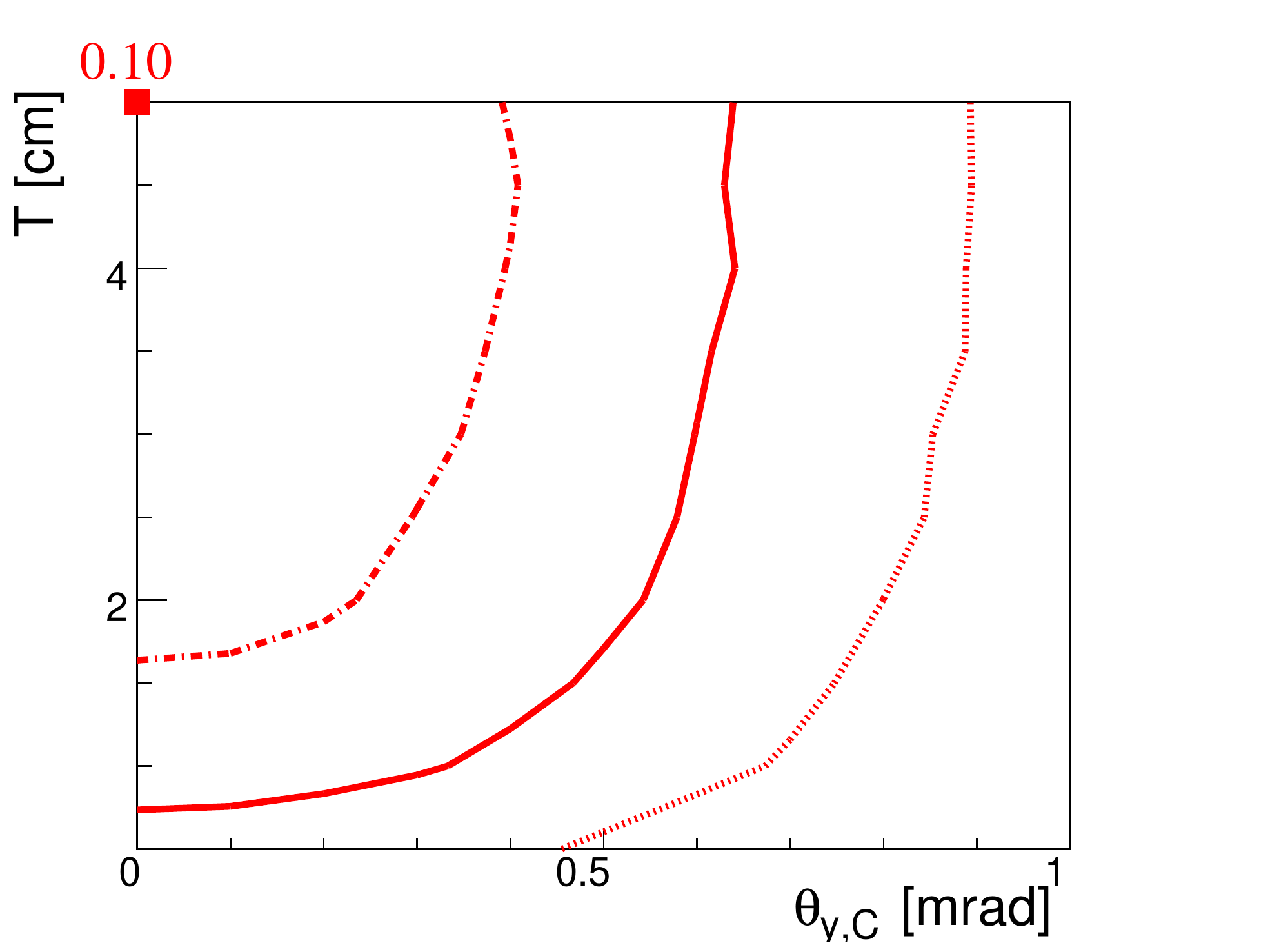}} \\
\subfigure[]{\includegraphics[width=0.75\columnwidth]{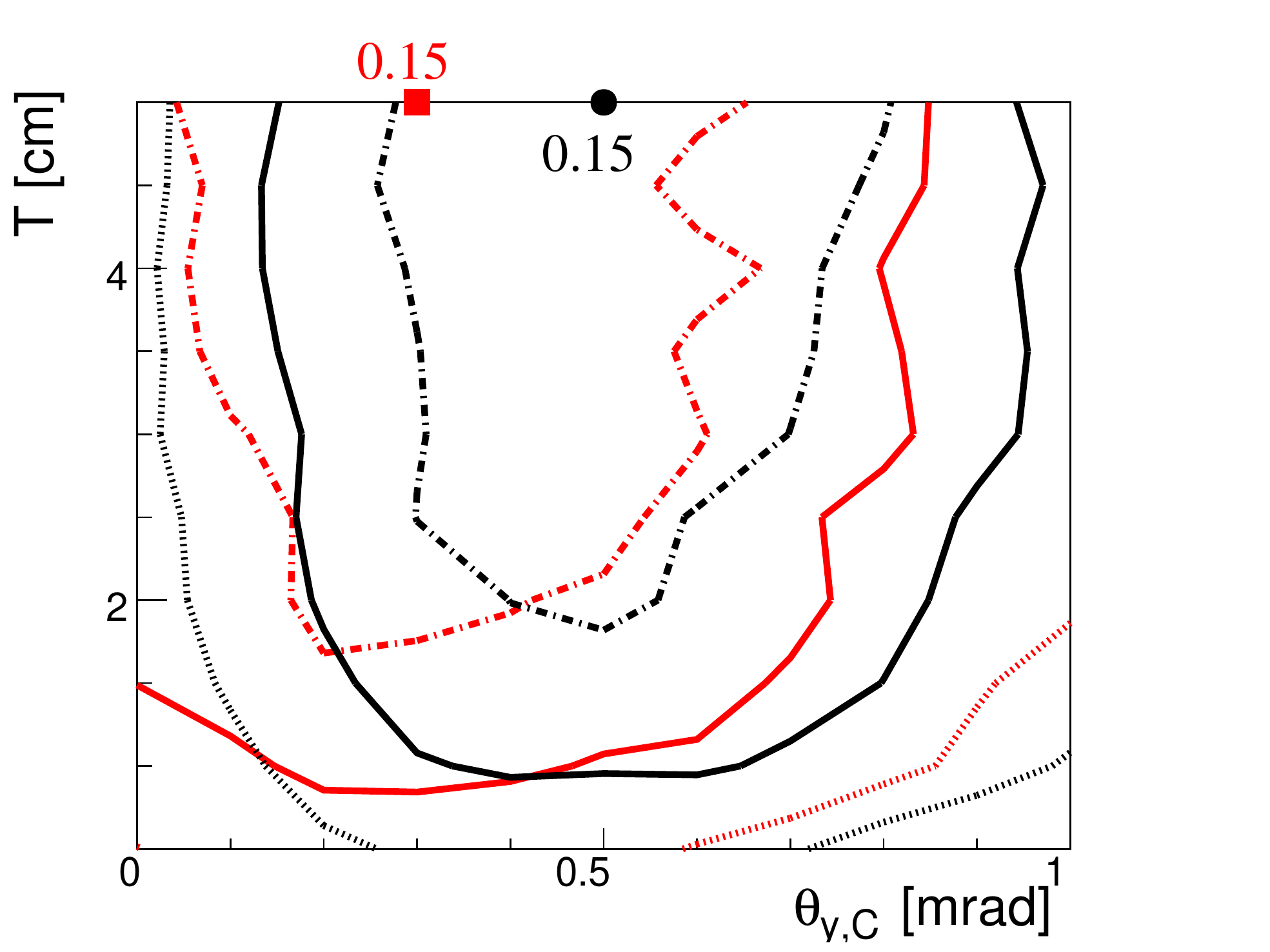}}
\caption{Regions of miniminal uncertainty of the $g$ (red curves) and $d$ (black) factors as a function of the crystal orientation angle $\theta_{y,C}$ and the target
thickness $T$, for (a) MDM measurement alone and
(b) combined MDM and EDM measurement, with magnetic anomalous moment $a \approx-0.03$ and $\Lc\to\pr\Km\pip$ decays.
The markers and values represent the minimum
uncertainty on the $g$ and $d$ factors
relative to $1.37\times 10^{13}$ \pot, corresponding to two years of
data taking at a rate of $10^6~\ppsec$~\cite{Mirarchi:2019vqi},
using a 16\mrad bent, 10\cm long germanium crystal at room temperature.
The curves are the regions
whose uncertainties are increased by 20\%, 50\% and 100\% with respect to the minimum.
A similar behaviour is observed for other considered crystal configurations.
\label{fig:optimization_thyCvsT}
}
\end{center}
\end{figure}

Following this approach, an enhanced sensitivity is obtained
while retaining the complete dependence on $g$, $d$ and $\gamma$, reducing significantly the need of a large crystal orientation angle $\theta_{y,C}$
for the EDM case~\cite{Fomin:2019wuw}, and solving discrete
ambiguities in the simultaneous extraction 
of the dipole moments~\cite{Bagli:2017foe}.
Pseudoexperiments have been generated assuming a proton flux on
target of about $10^{6} ~\proton/\sec$, which corresponds to $4.3\times 10^{10}$ protons on target (\pot) in ten hours operations of the \lhc~\cite{Mirarchi:2019vqi},
about $1.37\times 10^{13}$ \pot can be delivered by the accelerator in two years with an operational efficiency of about 55\%.
In Fig.~\ref{fig:optimization_thyCvsT}(a) the optimal
configuration for the measurement of the MDM
alone, \ie using Eq.~(\ref{eq:eom_sxsy_full}) with $d=0$,
is obtained at $\theta_{y,C}=0$,  which corresponds to the maximum
initial polarization along the $y$ axis, perpendicular to the
magnetic field ${\boldsymbol B^*}$.
In Fig.~\ref{fig:optimization_thyCvsT}(b), the optimal configuration
 for the combined measurement of MDM and EDM, \ie using Eq.~(\ref{eq:eom_sxsy_full}) with $d$ as free parameter, is obtained at $\theta_{y,C}$
 different from zero.
For the MDM the difference with respect to Fig.~\ref{fig:optimization_thyCvsT}(a) is due to limited \pot sample and to correlations between
the $a$ and $d$ parameters. 
The sensitivity to EDM requires
 a polarization component perpendicular to the electric
 field ${\boldsymbol E^*}$, \eg initial polarization along the $x$ axis.
However, as illustrated in Fig.~\ref{fig:optimization_thyCvsT}(b), for small $a \approx -0.03$~\cite{Dainese:2019xrz},
the decrease of EDM sensitivity with $\theta_{y,C} \approx 0.3 \ (0) \mrad$ with respect to the maximal sensitivity at $0.5 \mrad$
amounts to about 20\% (100\%), with marginal impact on the MDM (for further details see Appendix~\ref{sec:methods_sensitivity}).

A smaller crystal orientation angle reduces potential systematic uncertainties due to initial
polarization along the crystal $x$ axis, mimicking EDM effects.
Nevertheless, simulation studies show that a precise determination of the polarization model allows such systematic uncertainty to be kept under control.
The study of charm baryons with opposite polarization 
and the use of up- and down-bending crystals,
inducing opposite spin precession, 
offer complementary tools to prove the robustness of the results and
control systematic uncertainties~\cite{PhysRevAccelBeams.22.081004}.
For large $a \approx -0.76$~\cite{Dainese:2019xrz} the choice of crystal orientation has no impact on the sensitivities.

In addition, in Fig.~\ref{fig:optimization_thyCvsT} the \W target thickess is optimized considering the inelastic interactions of charm baryons
within the target, regulated by the nuclear interaction length and the
survival probability due to its lifetime~\cite{Tanabashi:2018oca} (for details see Appendix~\ref{sec:methods_aftertarget}).
A \W target thickness of $2\cm$,
about a factor of two thinner than that where maximal \Lc and \Xicp yields at the target exit are reached,
is considered as operational value.
With this choice the charm baryon dipole moment sensitivies are reduced by about 20\% with respect to the optimal thickness reached around 5 \cm,
as illustrated in Fig.~\ref{fig:optimization_thyCvsT} for \Lc baryons,
while detector occupancies are largely mitigated,
as required for detector operation and safety.

\begin{table}[htb!]
{\begin{tabular}{lcccc}
\hline
\rule{0pt}{2.4ex}$S^2$                                               & \multicolumn{4}{c}{$\approx 0.145$}            \\
$\epsilon_\textrm{DET}\ (\%)$                                          & \multicolumn{4}{c}{$0.20$}                      \\
Crystal configuration                                                & Ge    & Si    & Ge $77\degk$  & Ge $\Sb$ \\ 
\phantom{Crystal} angle (\!\mrad)                      &  16    & 16    & 16  & 7 \\
\phantom{Crystal} length (\!\cm)                      & 10      & 10    & 10  & 7 \\ 
\hline
\multicolumn{5}{c}{\rule{0pt}{2.4ex} \Lc baryon}   \\
\rule{0pt}{2ex}$\sigma$ (\mub/nucleon)                               & \multicolumn{4}{c}{$10.6$} \\
$a=(g-2)/2$                                                          & \multicolumn{4}{c}{$\approx -0.03~[-0.76]$} \\
$d$                                                    & \multicolumn{4}{c}{$0$} \\
$\br_{\textrm{eff}}\ (\%)$                                               & $20.2$        &  $19.5$      &  $20.6$    &  $20.6$         \\
$\epsilon_\textrm{CH}\ (\times 10^{-4})$                                &  $4.5$        &  $2.7$       &  $8.7$      &  $11.1$         \\
$\epsilon_\textrm{DF}\ (\times 10^{-2})$                                &  $3.3$        &  $1.7$       &  $4.9$      &  $13.1$         \\
$N_{\textrm{rec}}$                                                       &  $586$        &  $181$       &  $1748$     &  $5879$        \\
$\langle \gamma \rangle$                                              &  $709$        &  $573$       &  $834$      &  $855$          \\
$\langle \pt \rangle\ (\gevcc)$                                       & $0.79$        &  $0.71$      & $0.86$      & $0.87$         \\
$s_x\ (\%)$                                                           &  $11.8$       &  $8.6$       &  $14.1$     &  $15.6$        \\
$s_y\ (\%)$                                                           &  $-15.3$      &  $-14.2$     &  $-16.1$    &  $-16.1$        \\
$\sigma_{\boldsymbol \mu}\ (\times 10^{-2}\ {\boldsymbol \mu_{\bf N}})$      & $1.6$         & $3.4$        &  $0.8$      &  $0.9$          \\
$\sigma_{\boldsymbol \delta}\ (\times 10^{-16}\ e\cm)$                      & $2.2~[9.8]$   & $5.6~[17.1]$ & $0.9~[5.7]$ & $1.0~[2.9]$    \\
\hline
\multicolumn{5}{c}{\rule{0pt}{2.4ex} \Xicp baryon}   \\
\rule{0pt}{2ex}$\sigma$ (\mub/nucleon)                                & \multicolumn{4}{c}{$7.5$} \\
$a=(g-2)/2$                                                           & \multicolumn{4}{c}{$\approx 0.05~{[-0.47]}$} \\
$d$                                                    & \multicolumn{4}{c}{$0$} \\
$\br_{\textrm{eff}}\ (\%)$                                               & $6.1$        &  $5.8$      & $6.2$        & $6.2$  \\
$\epsilon_\textrm{CH}\ (\times 10^{-4})$                                & $5.7$        &  $3.8$      & $10.4$       & $12.2$  \\
$\epsilon_\textrm{DF}\ (\times 10^{-2})$                                & $10.7$       &  $7.6$      & $13.0$       & $24.6$ \\
$N_{\textrm{rec}}$                                                       & $627$        &  $284$      & $1429$       & $3146$ \\
$\langle \gamma \rangle$                                              & $514$        &  $433$      & $576$        & $588$ \\
$\langle \pt \rangle\ (\gevcc)$                                       & $0.74$       &  $0.69$     & $0.78$       & $0.79$ \\
$s_x\ (\%)$                                                           & $8.6$        &  $6.7$      & $10.2$       & $10.5$  \\
$s_y\ (\%)$                                                           & $-15.7$      &  $-14.8$    & $-16.3$      & $-16.5$ \\
$\sigma_{\boldsymbol \mu}\ (\times 10^{-2}\ {\boldsymbol \mu_{\bf N}})$      & $1.8$        & $3.1$       & $1.0$        & $1.5$ \\
$\sigma_{\boldsymbol \delta}\ (\times 10^{-16}\ e\cm)$                      & $3.0~[5.1]$  & $5.9~[6.8]$ & $1.5~[3.5]$  & $2.4~[2.1]$ \\
\hline
\end{tabular}}
\caption{Relevant parameters for \Lc and \Xicp baryons considered
for sensitivity studies:
the average event information $S^2$  (assumed to be similar for all modes),
the detector efficiency $\epsilon_\textrm{DET}$, which includes trigger, reconstruction and selection criteria of signal events based upon the $3h$ system,
the baryon production cross section $\sigma$,
the anomalous magnetic moment $a$ (two values are used from Ref.~\cite{Dainese:2019xrz}, the second is indicated in squared brackets),
the gyroelectric factor $d=0$,
the effective branching ratio $\br_{\textrm{eff}}$,
the channeling efficiency $\epsilon_\textrm{CH}$, which includes the efficiency of the particle to be trapped into channeling regime,
the decay flight efficiency $\epsilon_\textrm{DF}$ of the baryon within the crystal length,
the number of reconstructed charm baryons $N_{\textrm{rec}}$,
the average boost $\langle \gamma \rangle$ and transverse momentum $\langle \pt \rangle$,
the initial spin-polarization $s_x$,
and the initial spin-polarization $s_y$ for events with positive $p_{x_L}$ ($-s_y$ for negative $p_{x_L}$).
Uncertainties on the MDM and EDM,
$\sigma_{\boldsymbol \mu}$ and $\sigma_{\boldsymbol \delta}$, respectively,
are relative to $1.37\times 10^{13}$ \pot.
At \lhcb, silicon and germanium bent crystals
at room temperature, and germanium at $77\degk$, are considered. For comparison, the \Sb scenario is evaluated with germanium at room temperature.
The corresponding crystal parameters are reported in the table.
The target thickness is 2 \cm in all cases.
All three- and four-body modes from Tables~\ref{tab:Lc_decay_modes} and~\ref{tab:Xic_decay_modes} are considered.
}
\label{tab:sensitivity}
\end{table}

\begin{figure}[htb!]
{\begin{tabular}{cc}
\subfigure[]{\includegraphics[width=0.47\columnwidth]{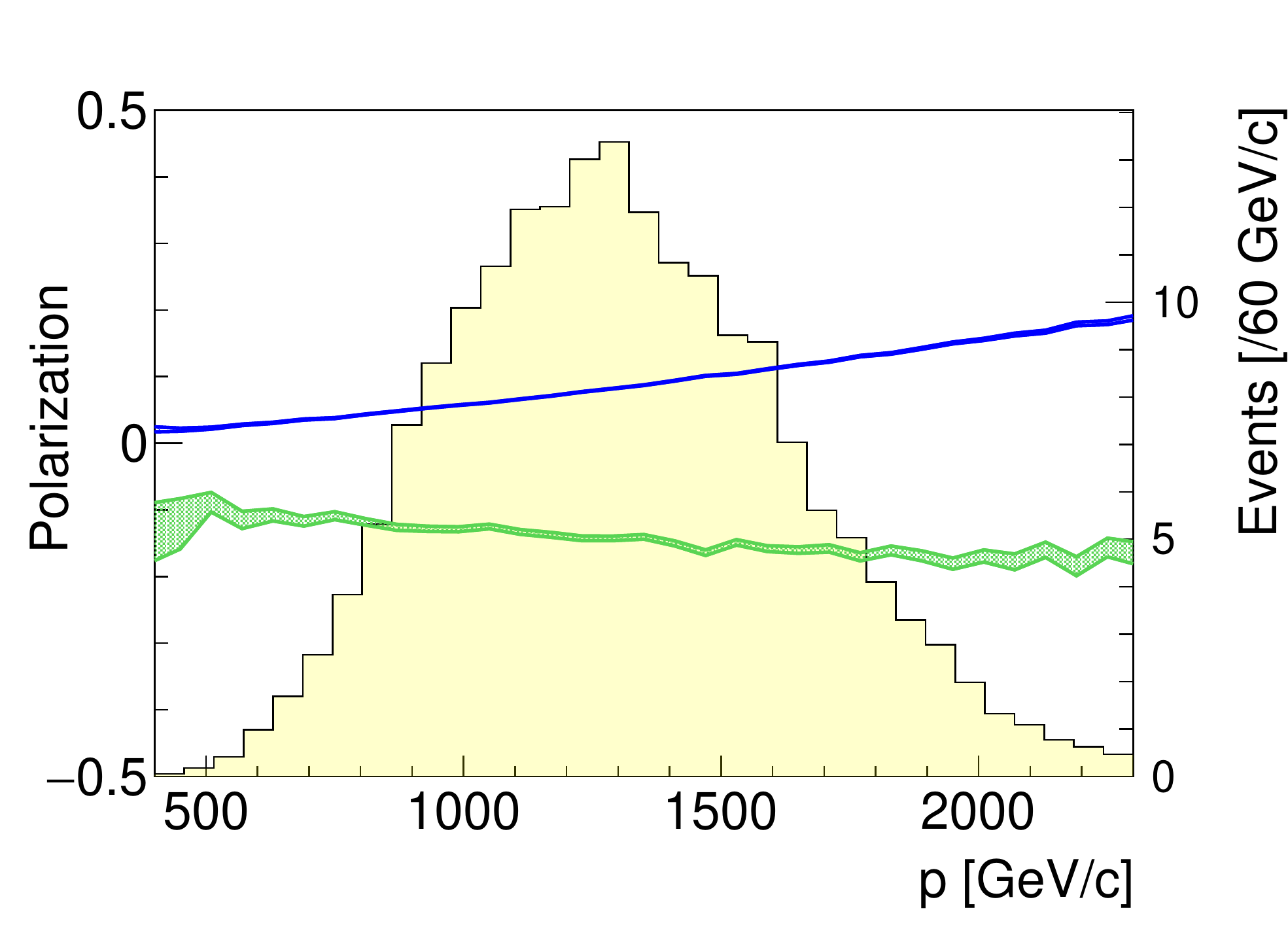}} &
\subfigure[]{\includegraphics[width=0.47\columnwidth]{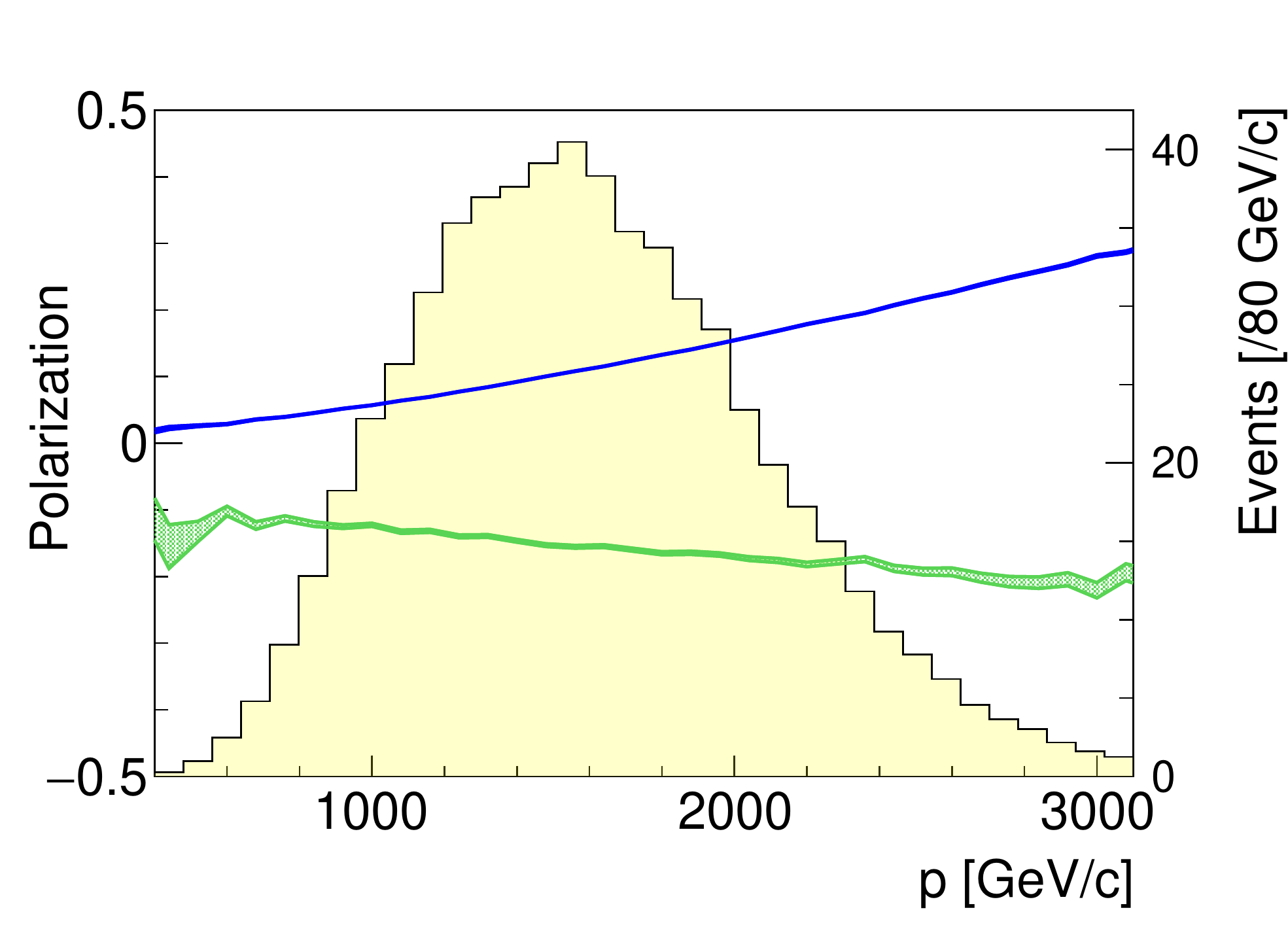}} \\
\subfigure[]{\includegraphics[width=0.47\columnwidth]{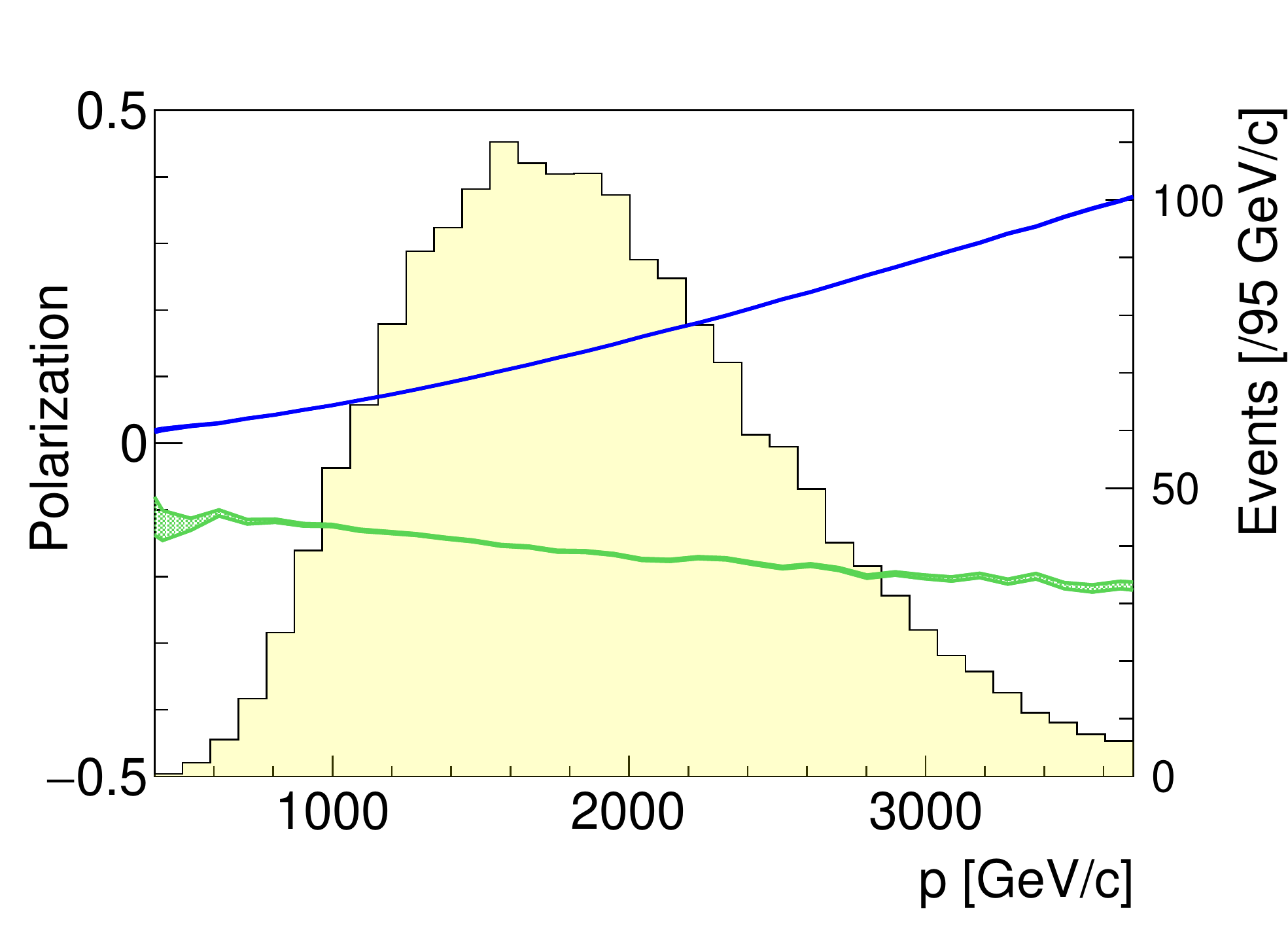}} &
\subfigure[]{\includegraphics[width=0.47\columnwidth]{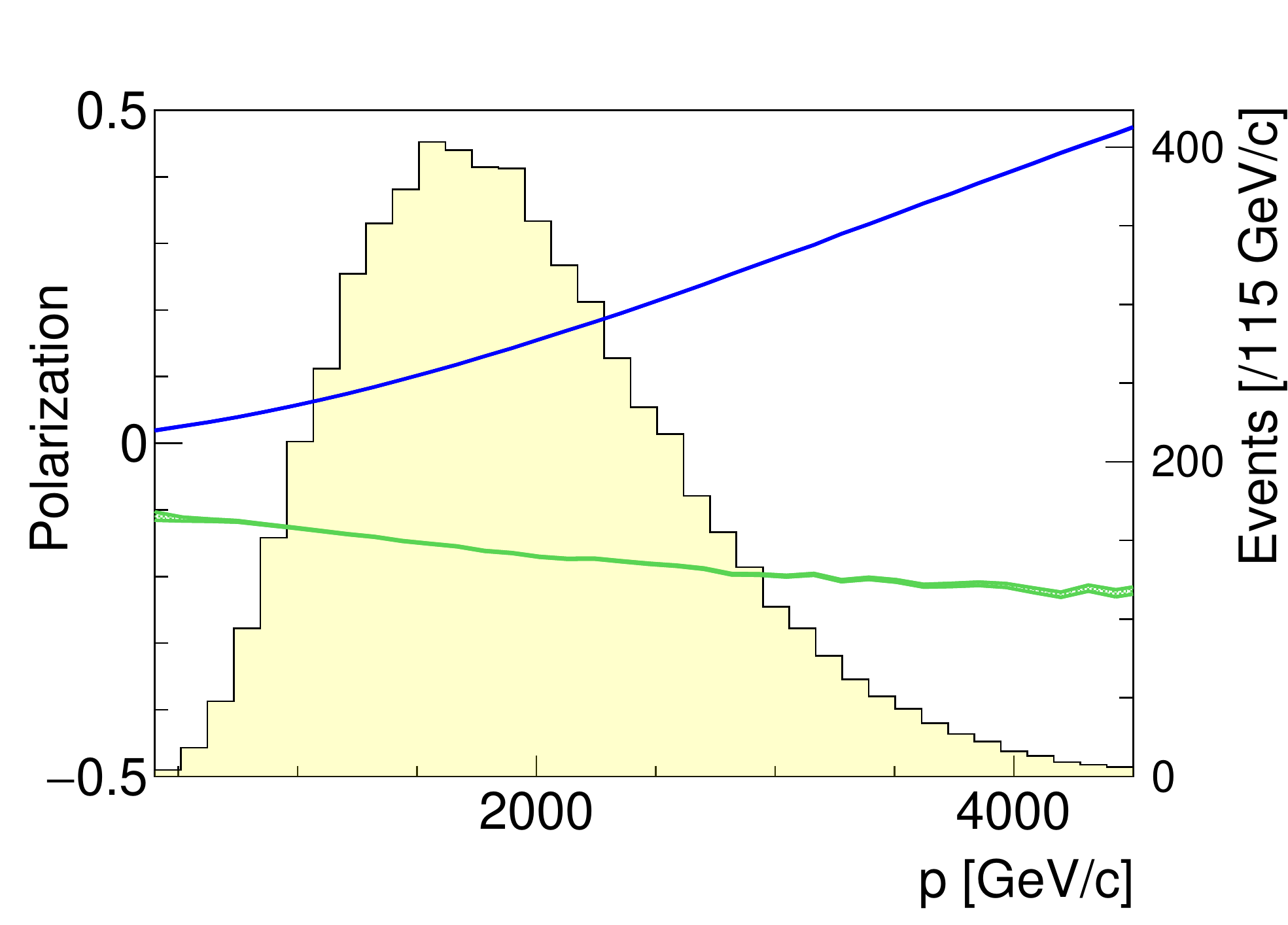}}
\end{tabular}}
\caption{\label{fig:sensitivity_spectra_Lc}
  Initial polarizations $s_x$ (hatched blue bands) and $s_y$ for events with positive $p_{x_L}$ ($-s_y$ for negative $p_{x_L}$, solid green) as a
  function of the \Lc baryon momentum
  for (a) silicon $293\degk$, (b) germanium $293\degk$, (c) germanium $77\degk$, and (d) germanium $293\degk$ for \Sb, with parameters reported in the text.
  The bands represent one standard deviation regions from the pseudoexperiments. 
  The histograms, normalized to $1.37\times10^{13}$~\pot, show the 
  spectra of channeled and reconstructed particles.
}
\end{figure}
\begin{figure}[htb!]
{\begin{tabular}{cc}
\subfigure[]{\includegraphics[width=0.47\columnwidth]{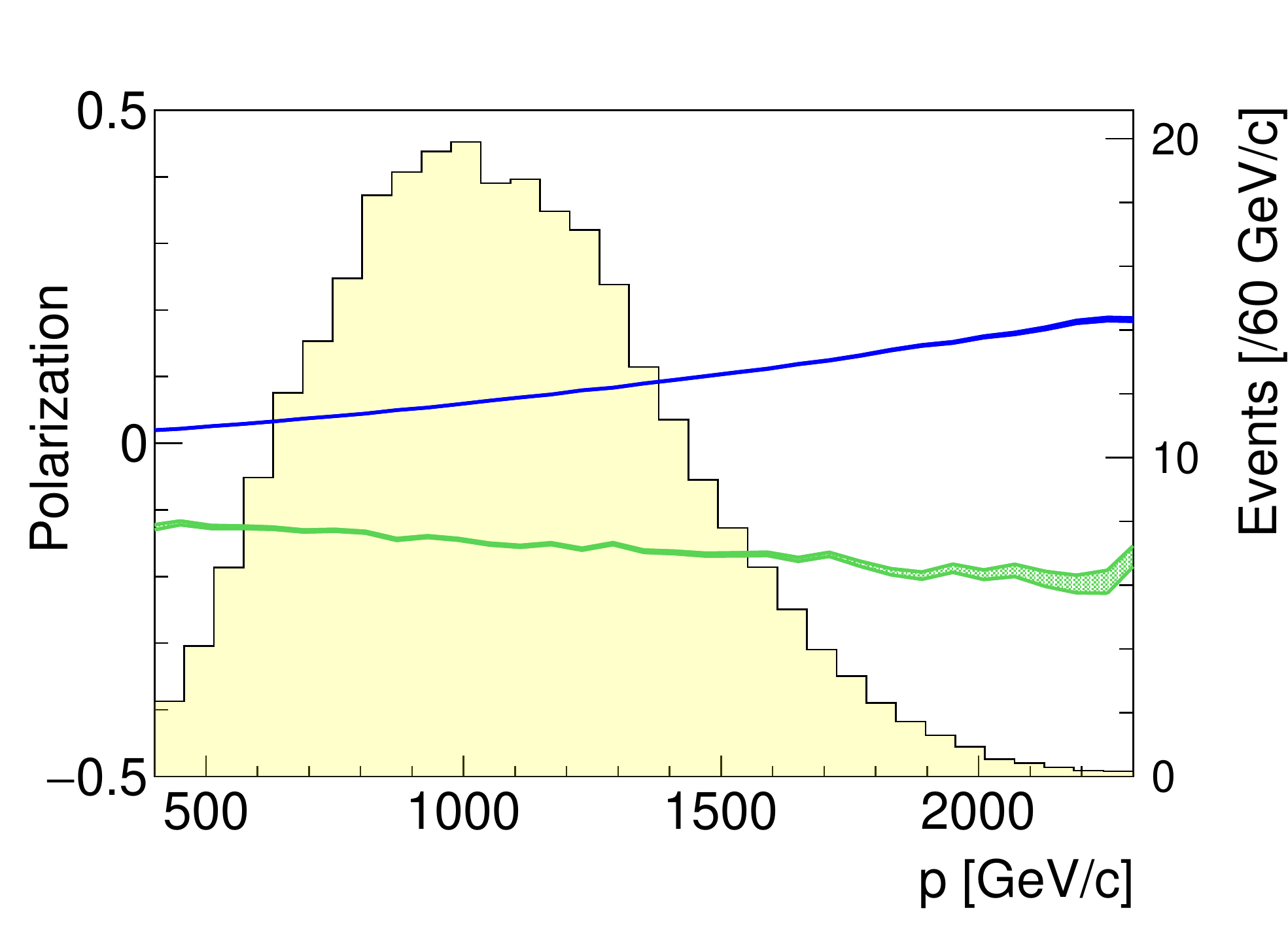}} &
\subfigure[]{\includegraphics[width=0.47\columnwidth]{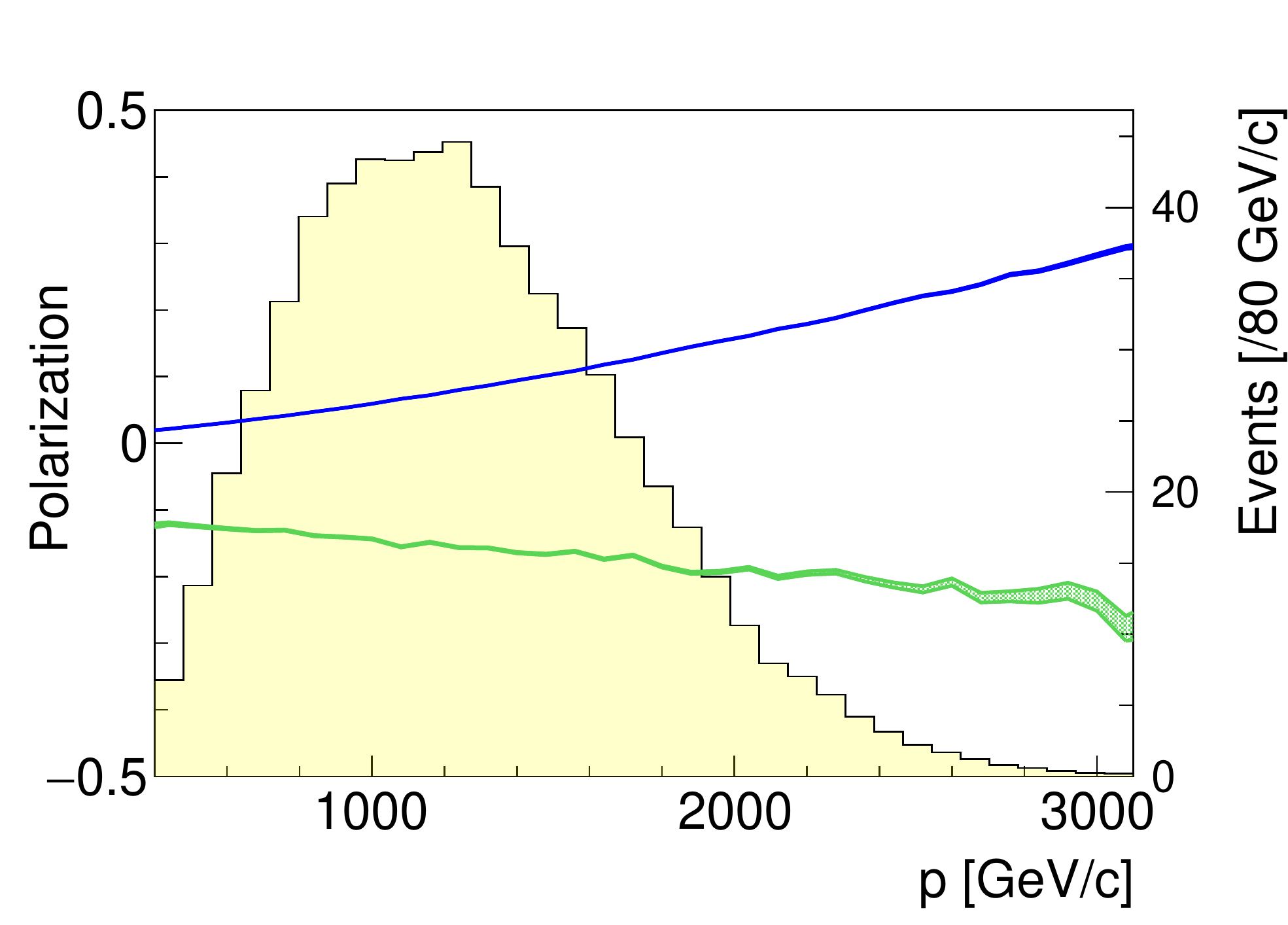}} \\
\subfigure[]{\includegraphics[width=0.47\columnwidth]{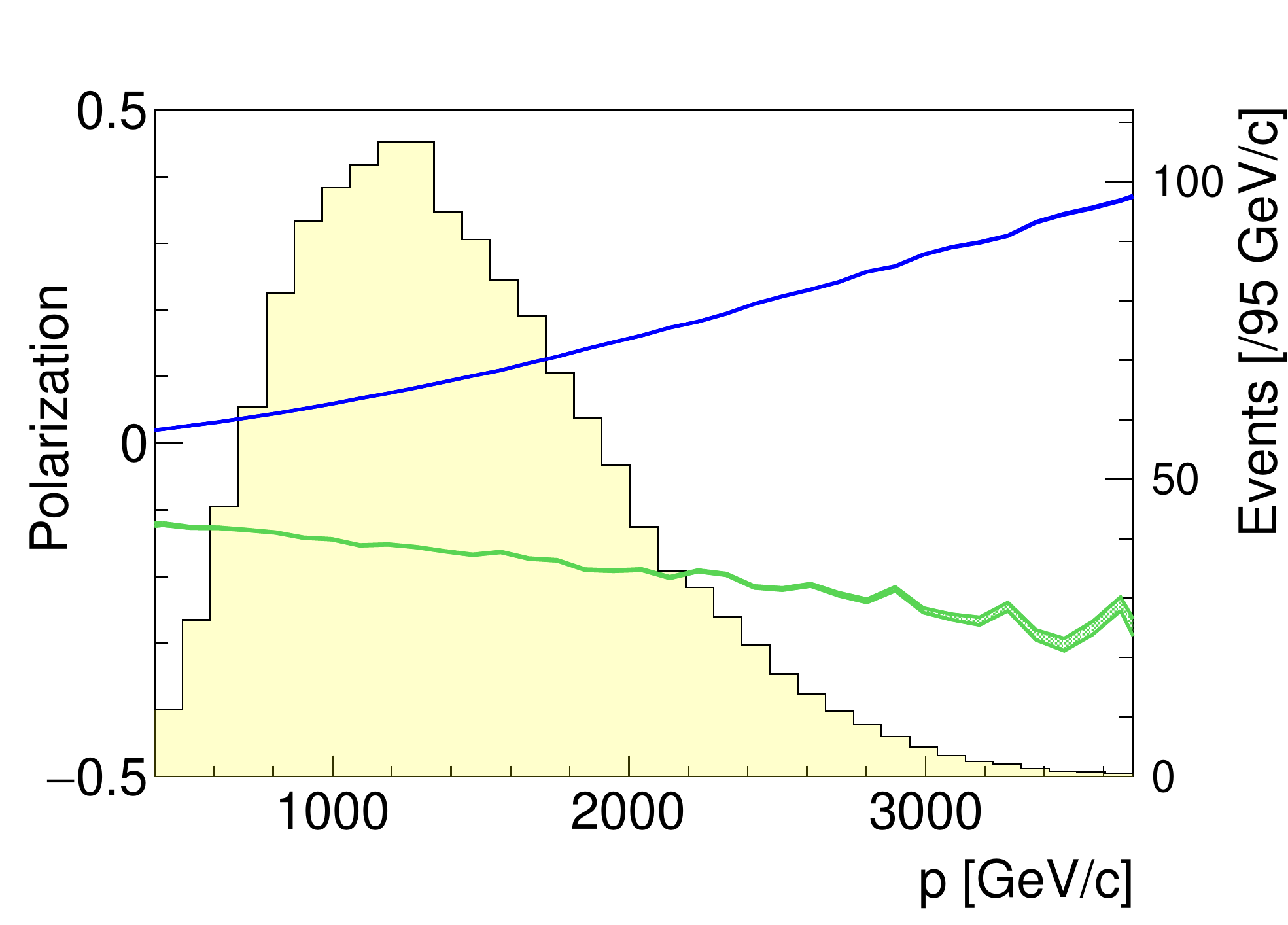}} &
\subfigure[]{\includegraphics[width=0.47\columnwidth]{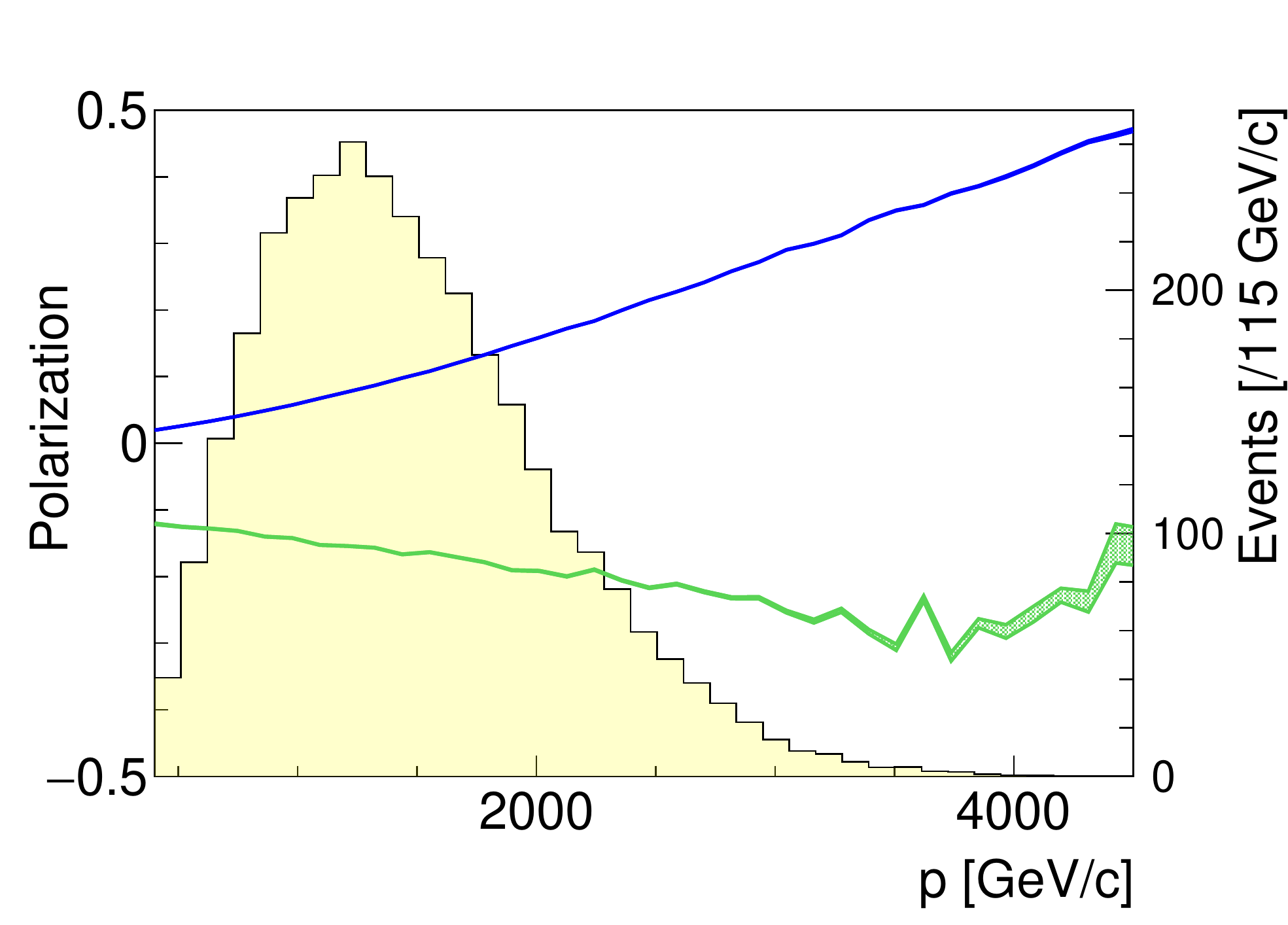}}
\end{tabular}}
\caption{\label{fig:sensitivity_spectra_Xc}
Same as Fig.~\ref{fig:sensitivity_spectra_Lc} for \Xicp baryons. The oscillations in the initial polarization bands (particularly visible for $s_y$) are
due to the limited sample size of the simulations.
}
\end{figure}

\begin{figure}[htb!]
\begin{center}
\subfigure[]{\includegraphics[width=0.90\columnwidth]{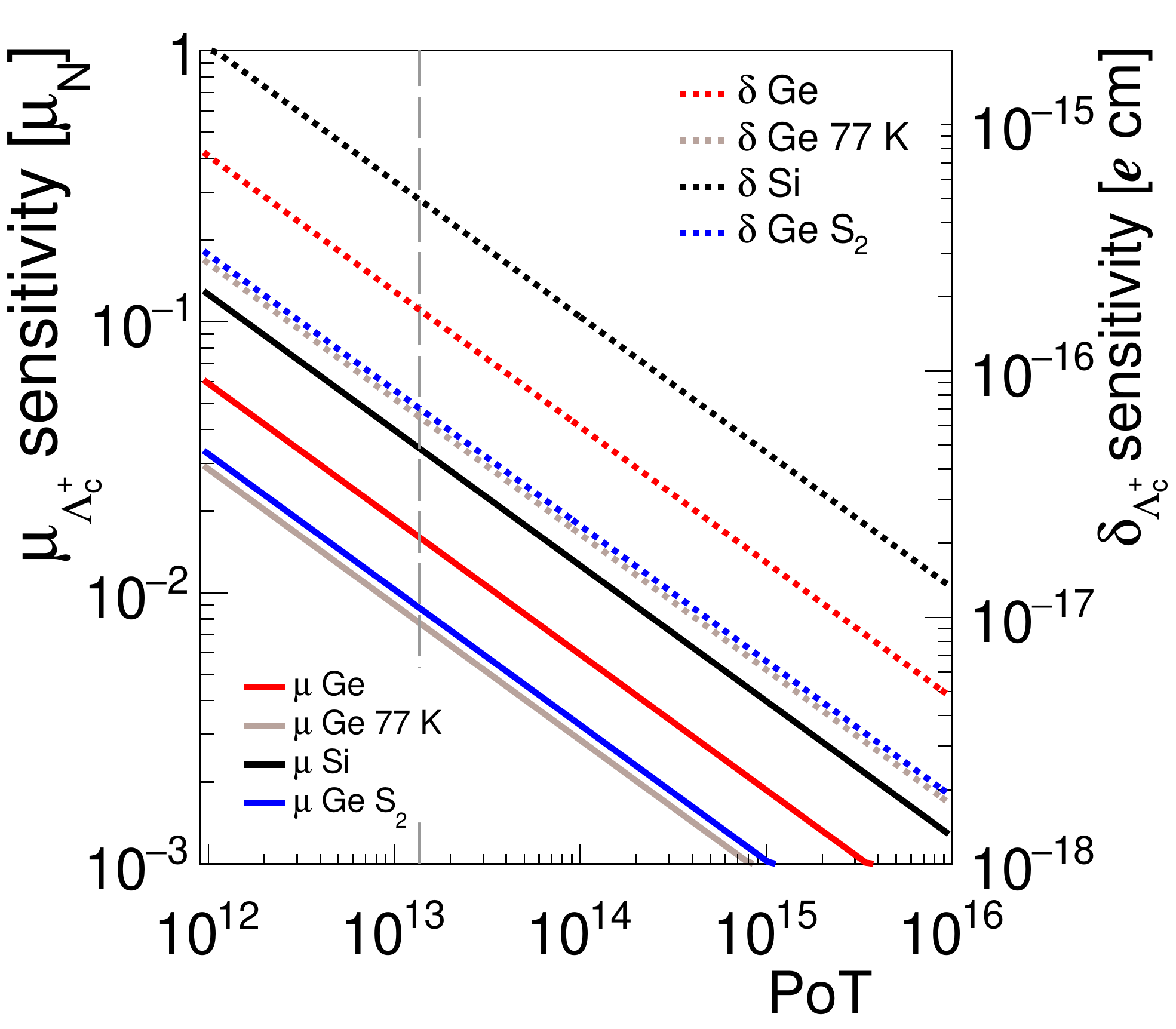}} \\
\subfigure[]{\includegraphics[width=0.90\columnwidth]{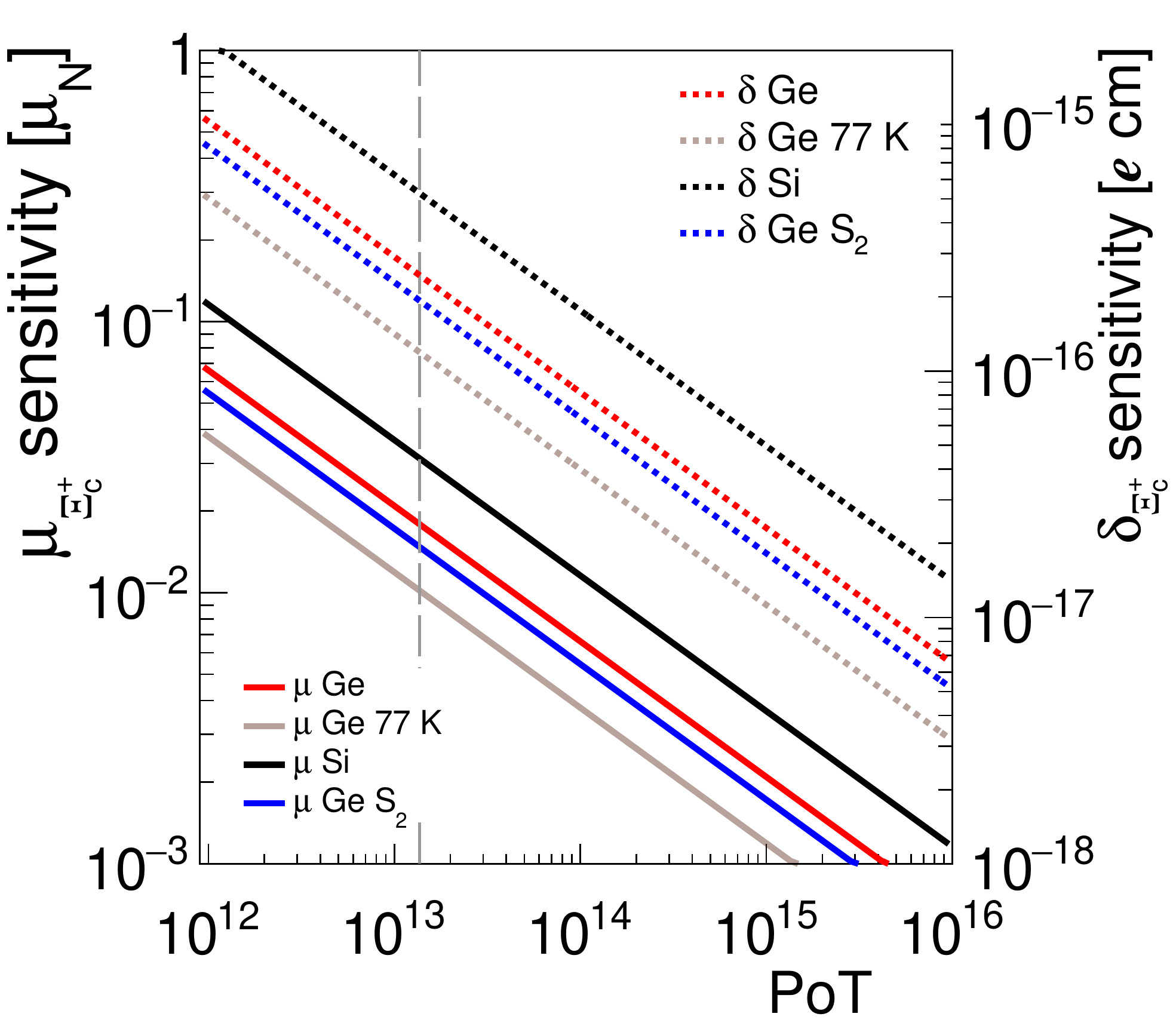}}
\caption{\label{fig:sensitivity_LcXc} Uncertainties on the MDM and EDM of \Lc (a) and \Xicp (b) baryons
as a function of \pot at \lhcb and at a dedicated experiment (\Sb) with increased forward acceptance.
The sensitivities at \lhcb for
room temperature $293\degk$ silicon and germanium bent crystals are
compared with a germanium crystal cooled at $77\degk$, with parameters reported in the text.
All three- and four-body \Lc and \Xicp decays from Tables~\ref{tab:Lc_decay_modes} and~\ref{tab:Xic_decay_modes} are considered,
with anomalous magnetic moment $a$ assumed to be $\approx -0.03$ and $\approx 0.05$, respectively.
The vertical long-dashed lines refer to $1.37\times10^{13}$~\pot, corresponding to two years of data taking.
}
\end{center}
\end{figure}

Relevant parameters for the sensitivity studies are reported in Table~\ref{tab:sensitivity},
along with the expected signal yields and uncertainties on the MDM and EDM in two years of data taking with the \lhcb detector, for different crystal configurations.
Figures~\ref{fig:sensitivity_spectra_Lc} and~\ref{fig:sensitivity_spectra_Xc} show the corresponding reconstructed \Lc and \Xicp momentum spectra,
along with the estimated $s_x$ and $s_y$ initial polarizations as a function of the baryon momentum.
The latter are determined by the convolution of 
Eq.~(\ref{eq:pol_init}) with the transverse momentum distribution of channeled baryons. 
How the sensitivities evolve as a function of \pot is illustrated in Fig.~\ref{fig:sensitivity_LcXc}.
Silicon and germanium crystals with deflection angle of 16\mrad and 10\cm length are considered for compatibility
with the \lhcb detector acceptance and operations, while maximizing experimental sensitivity.
For the case of germanium, sensitivity results at $77\degk$ in addition
to room temperature are shown.
The scenario of a future dedicated experiment at the \lhc,
indicated as \Sb, 
is also studied using a setup based on a germanium crystal at room temperature
of $7\mrad$ bending angle and $7\cm$ length, which would allow
effective separation of channeled charm baryons from forwardly
produced background particles, to be kept outside the detector acceptance.
Identical detector performance and signal
reconstruction efficiency to the \lhcb apparatus are assumed,
but extended in the forward direction.
Nevertheless, in the \Sb scenario channeled baryons are deflected at smaller angles with high momentum, extending up to about 4 \tev,
and would require an advanced detector design with a long lever arm and intense magnetic field for precise
particle momentum measurements, and high granularity to be able to operate and reconstruct the events under high background levels in the very forward region.

Compared to the configuration of germanium at room temperature, the significantly higher yields with germanium \Sb, close to a factor of ten (five)
for \Lc (\Xicp) baryons, and the 20\% (15\%) harder momentum spectrum, reflect in an increase of MDM sensitivity equivalent to a data
sample about three times larger (same size). This is a consequence of the different bending angles at \lhcb and \Sb, \ie 16 and 7 \mrad, respectively.
For germanium at $77\degk$, with yields higher by a factor of three (two) and momentum spectra similar
to germanium \Sb, the uncertainties are reduced by an equivalent data sample close to a factor of four (three) larger.
Silicon provides significantly lower yields and softer momentum spectra, resulting in sensitivities whose
equivalent data sample size is a factor of four (three) smaller.
EDM sensitivities strongly depend on the anomalous magnetic moment $a$, although they present similar features. 
Thus, improved sensitivities with respect to
the \lhcb detector based configurations
are expected at \Sb if much higher \pot would be achievable.

%
%

%
\section{Conclusions}
In summary, progress towards the first measurement of charm baryons dipole
moments is reported. 
An experimental setup based on bent crystals and a \W target
placed upstream of the \lhcb detector is studied.
Silicon and germanium long bent crystals have been
tested on a $180~\gev$ hadron beam with relatively high channeling
efficiency measured for both prototypes.
The germanium crystal provides enhanced  sensitivity to MDM and EDM compared
to the silicon crystal, especially when cooled down at $77\degk$~\cite{Bezshyyko:2017var}. 
Advanced analysis techniques have been developed for three- and four-body
charm baryon decays,
providing enhanced sensitivity to the measurements.
For a baseline configuration with $1.37\times 10^{13}$ \pot impinging
on 2 \cm~\W target and with a germanium crystal at room temperature,
similar sensitivities for the MDM (EDM) of \Lc and
\Xicp baryons below $2\times 10^{-2}\ {\boldsymbol \mu_{\bf N}}$ ($ 3\times 10^{-16}~e\cm$) are achievable, which corresponds to a relative precision on the gyromagnetic
factors $g$ of about 4\%.
A germanium crystal cooled at 77 \degk would improve the sensitivity by a factor of two.
These results enable a unique program of MDM and EDM measurements
of charm baryons at \lhcb,
capable to test advanced
low energy models of strong interactions
and to search for physics beyond the SM.
A future dedicated experiment with significantly higher \pot
would offer the possibility to improve the sensitivity to charm baryon dipole
moments, and to explore beauty baryons and ultimately the $\tau$ lepton~\cite{Fomin:2018ybj,Fu:2019utm}.
%
%
\appendix
\hfill \break
\section{Manufacturing of crystal and its bender}
\label{sec:methods_manifacturing}

Silicon and germanium crystals are prepared starting from wafers available
from commercial suppliers. The thickness of the wafers are $5\mm$ and $1\mm$
for the silicon and germanium wafers, respectively.
Wafers with a dislocation density lower
than $1/\cm^2$ over the entire region interacting with the particle beam are
selected from a stock of wafers. The density of dislocations is characterized
through the etch pit density~\cite{Sirtl:1961,Aragona:1972} and the X-ray
topography techniques.
The miscut angle, \ie the angle between the optical surface of the wafer surface
and the atomic planes, is measured using a high-resolution X-ray
diffractometer (Panalytical X'Pert$^3$ MRD XL) coupled to an autocollimator.
Subsequently, the miscut is reduced to less than $0.01^\circ$ through polishing
with a Logitech PM5 equipment. The wafers are diced with a dicing machine
(Disco DAD3220) to rectangular crystals, then cleaned in hot bath of
acetone under ultrasonic agitation.
Bending of the crystal occurs as consequence of clamping of the crystal
between surfaces of a properly machined bender. Assemblies describing crystal
benders are modeled through finite element modelling (Ansys R18), and the
shape of the surface in contact with the crystal are properly modelled to
maximize uniformity of the deformation of the crystal.
Bender are manufactured through milling and electro discharge machining
of a block of stainless steel 316LN. This material are chosen for its
compatibility with the environment of \lhc, where the devices are supposed to
operate. After machining, the bender is cleaned in acetone bath under
ultrasonic agitation.
To avoid interference of dust which might deposit on the surfaces of the
bender or of the crystal, the assembly of crystal on the bender is accomplished
in a ISO-4 clean room.
\section{Simulation studies}
\subsection{Simulation of interaction between particle beam and crystal}
\label{sec:methods_simuinteractions}

\begin{figure}[h]
  \begin{center}
\includegraphics[width=0.90\columnwidth]{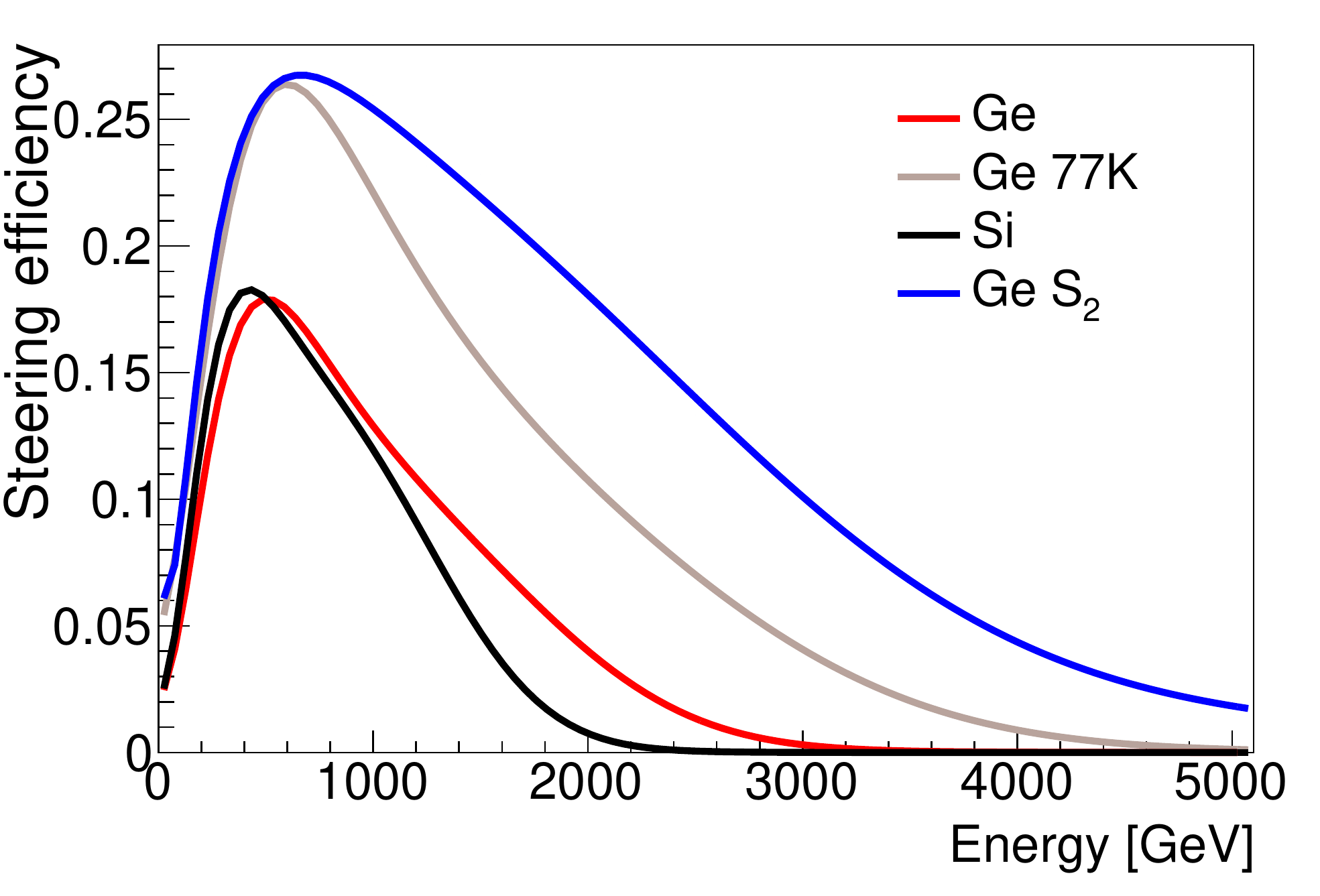}
\caption{\label{fig:crystals_simulations}
  Monte Carlo simulation results for the steering efficiency vs energy of particles impinging parallel to
the atomic planes, assuming uniform bending radius,
  of germanium and silicon crystals with deflection
  angle 16\mrad and 10\cm length (\lhcb scenario), and germanium 7\mrad bent and 7\cm length ($S_2$), at room temperature. For germanium in the
  \lhcb scenario results with cooling at $77\degk$ are also shown.
}
\end{center}
\end{figure}

A precise treatment of scattering of channeled particles in the crystal is fundamental
for the design of experiments aiming to study spin precession of short lifetime particles.
Scattering from nuclei and inner shell electrons play an important role in describing the dynamics
of channeled particles, rendering the description of this process more difficult.
Propagation of particles in the crystal is described by the Monte Carlo simulation
CRYSTALRAD~\cite{Sytov:2014jfa,Sytov:2019gad}.
State of the art simulations are based on the solution of the equation of motion of a charged
particle interacting with a crystalline lattice and are quantum-mechanically grounded,
including refined treatments of both large and small-angle scattering.
Recently observed experimentally~\cite{Mazzolari:2019fus} incoherent scattering modification
is also part of modern simulation codes.
Bent crystals cooled to cryogenic temperatures are more efficient than
at room temperature due to the lower vibration amplitude of the
atoms~\cite{Forster:1989eh}.
According to channeling simulations the
improvement is more significant for germanium than for silicon crystals.
Steering efficiencies for different crystal configurations considered in the sensitivity studies, as obtained from these simulations,
are shown in Fig.~\ref{fig:crystals_simulations}.

\subsection{Evaluation of the $\boldsymbol \Lc$ and $\boldsymbol\Xicp$ spectra after the tungsten target and the crystal}
\label{sec:methods_aftertarget}

The variation per unit length of the number of protons, $N_\pr$, due to the interaction with the
tungsten target is given by
\begin{equation}
\label{eq:exp_int}
\frac{{\textrm d}N_p}{{\textrm d}z}=-\frac{N_p}{\lambda_\W} ,
\end{equation}
where $\lambda_\W = A_T / (\rho N_A  A_\W^\textrm{part} \sigma_{\pr N}) \approx 8.87\cm$ is the \W nuclear interaction length at $\sqrt{s} \approx 115\gev$,
with $\rho$ and $A_T$ the \W density and atomic mass
respectively, $N_A$ the Avogadro number,
$A_{\textrm W}^\textrm{part}$ the number of participant nucleons calculated on the basis of the Glauber model~\cite{Miller_2007,Loizides:2016djv},
and $\sigma_{\pr N}$ the proton-nucleon inelastic cross-section at $\sqrt{s} \approx 115\gev$~\cite{Tanabashi:2018oca}.

\begin{figure}[h]
\begin{center}
\includegraphics[width=0.90\columnwidth]{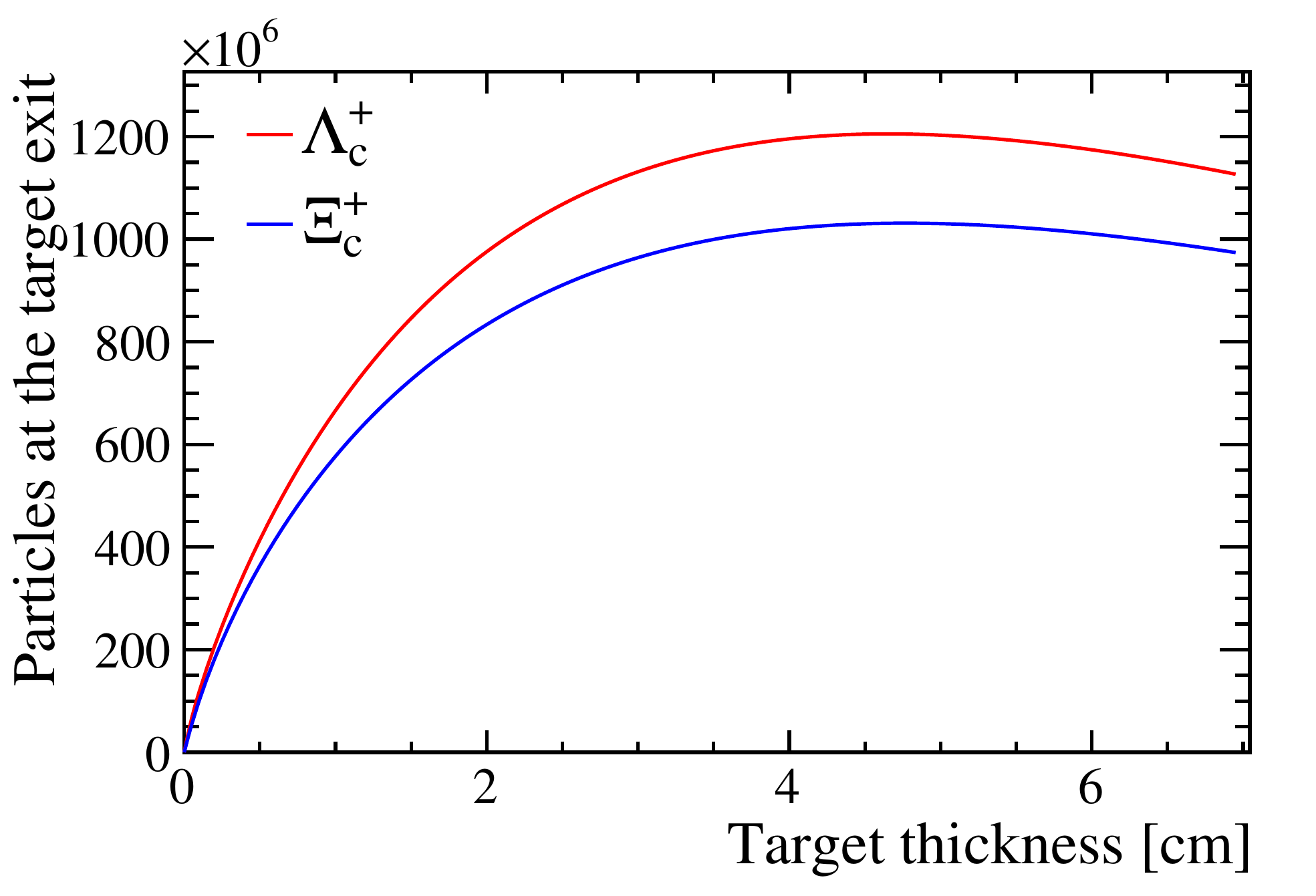}
\caption{\label{fig:nLc_afterTarget} Number of produced \Lc and \Xicp baryons exiting
 the \W target as a function of the target thickness. The case of 
 $1.37\times 10^{13}$ \pot, corresponding to two years running in \lhcb, is shown.}
\end{center}
\end{figure}

The variation per unit length of the number of \Lc (and similarly \Xicp) baryons
$N_\Lc$, is determined by the disappearance of protons
in the \pr\W interaction, and by the decay and nuclear interaction in the target of the produced \Lc baryons,
\begin{equation}
\label{eq:Lc_prod_diff}
\frac{{\textrm d}N_{\Lc}}{{\textrm d}z} = \frac{N_\pot}{\lambda_{\W,\Lc}}e^{-z/\lambda_\W}  - \frac{N_{\Lc}}{\lambda'},
\end{equation}  
where $1/\lambda'=1/\lambda_\W^{(\Lc)}+1/(\beta \gamma c \tau)$,
$\beta\gamma$ is the \Lc Lorentz boost factor, $\tau$ its lifetime, $c$ the speed of light,
$N_\pot$ the number of protons hitting the target,
\mbox{$\lambda_\W^{(\Lc)} \approx \lambda_\W$} the \Lc interaction length, and
$\lambda_{\W,\Lc}$ is the mean free path for \Lc production.
The latter is estimated as
$\lambda_{\W,\Lc} = A_T / (\rho N_A  A_N \sigma_{\Lc}) \approx 81.35\m$,
where $A_N$ is the \W atomic mass number
and $\sigma_{\Lc}$ the proton-nucleon to $\Lc$ cross-section at $\sqrt{s} \approx 115\gev$,
obtained by scaling linearly the measured proton-nucleon to \ccbar cross-section at $\sqrt{s} = 86.6\gev$~\cite{Aaij:2018ogq}
and using the \Lc fragmentation fraction~\cite{Lisovyi:2015uqa,Gladilin:2014tba}.
For \Xicp baryons the fragmentation fraction is estimated to be about $0.7$ times smaller than for \Lc,
assuming it to be similar for \Xicp, \Xicz and \Omegacz baryons~\cite{Bagli:2017foe},
leading to $\lambda_{\W,\Xicp} \approx 114.72\m$.
Integration of Eq.~(\ref{eq:Lc_prod_diff}) for a target of thickness $z$ and fixed $\beta\gamma$ yields
\begin{equation}
\label{eq:Nc_target}
N_{\Lc}(z,\beta\gamma) = \frac{N_\pot}{\lambda_{\W,\Lc}}\beta\gamma c\tau \left( e^{-z/\lambda_\W}  - e^{-z/\lambda'} \right).
\end{equation}
The number of \Lc baryons after traversing a thickness $z$ of the target, $N_{\Lc}(z)$, is then given by the convolution
of Eq.~(\ref{eq:Nc_target}) with the normalized \Lc baryon momentum spectra,
$r(\beta\gamma) = 1/\sigma_\Lc {\textrm d} \sigma_\Lc(\bp)/{\textrm d}^3\bp$,
estimated using Monte Carlo simulations based on \pythia.
Figure~\ref{fig:nLc_afterTarget} shows the number of \Lc and \Xicp baryons exiting the target as a function of the target thickness for the case of $N_\pot = 1.37\times 10^{13}$ protons.

The number of \Lc baryons produced in a target of thickness $T$ and traversing a bent crystal of length $L$ is given by the
convolution of
\begin{equation}
\label{eq:Nc_crystal}
N_{\Lc,\textrm{cry}}(T,L,\beta\gamma)= N_{\Lc}(T,\beta\gamma)\epsilon_\textrm{CH}(\beta\gamma)\epsilon_\textrm{DF}(L,\beta\gamma)
\end{equation}
with $r(\beta\gamma)$,
where $\epsilon_\textrm{CH}(\beta\gamma)$ is the steering efficiency of the crystal, which includes the efficiency of the \Lc particle
to be trapped into channeling regime,
and $\epsilon_\textrm{DF}(T,L,\beta\gamma) = e^{-L/\beta \gamma c \tau}$ accounts for the particle decay flight, \ie survival probability within the crystal length.
\subsection{Reconstruction of $\boldsymbol \Lc$ and $\boldsymbol \Xicp$ decays with a $\boldsymbol\piz$ in the final state}
\label{sec:methods_reco}

The momentum in the laboratory frame of a \Lc (and similarly for \Xicp) baryon
decaying into final states containing an undetected neutral pion can be
reconstructed, up to a two-fold ambiguity, using kinematic information of the remaining charged-particle decay products ($3h$).
The ambiguity arises from two different configurations in the
baryon rest frame which are indistinguishable in the laboratory frame. Either a relatively low momentum \Lc charm baryon decays into the $3h$ system with a small angle
with respect to the \Lc flight direction, or the $3h$ system is produced with a large angle from a higher momentum \Lc.
The ambiguity vanishes for events in which the angle $\theta$ between the flight direction of the baryon and the $3h$ system reaches its maximum allowed value,
\begin{equation}
\label{eq:sinthetamax}
\theta_\textrm{max} = \arcsin \frac{\sqrt{ \left( m^2 - m_{3h}^2 - m_{\piz}^2 \right)^2 - 4 m_{3h}^2  m_{\piz}^2 }}{2 m p_{3h}} ,
\end{equation}
leading to a single solution
\begin{equation}
\label{eq:momentumreco}
p = \frac{p_{3h} \left( m^2 + m_{3h}^2 - m_{\piz}^2 \right) \cos\theta_\textrm{max} }{2 \left( m_{3h}^2 + p_{3h}^2 \sin^2\theta_\textrm{max} \right)},
\end{equation}
which is used as estimate of the \Lc momentum magnitude.
Here, $p_{3h}$ and $m_{3h}$ are the momentum magnitude and invariant mass of the $3h$ system, respectively,
$m$ is the nominal charm baryon mass, and $m_{\piz}$ the neutral pion mass.
This procedure does not impose the additional kinematical contraint when the \piz meson is produced via \Sigmap decays. 
The \Lc three-momentum then can be determined from the \Lc flight direction, itself obtained from the knowledge of the
production (primary), $\boldsymbol x_\textrm{PV}$, and the \Lc decay, $\boldsymbol x_\Lc$, vertex positions, as $\bp = p \boldsymbol u$,
where $\boldsymbol u$ is a unit vector along the \Lc flight direction,
\begin{equation}
\label{eq:3momentumreco}
\boldsymbol u = \frac{\boldsymbol x_\Lc - \boldsymbol x_\textrm{PV}}{|\boldsymbol x_\Lc - \boldsymbol x_\textrm{PV}|}.
\end{equation}
The average initial spin-polarization vector can then be inferred from Eq.~(\ref{eq:pol_init}).

\begin{figure*}[htb!]
  \begin{center}
{\begin{tabular}{ccc}
\centering
\subfigure[]{\includegraphics[width=0.65\columnwidth]{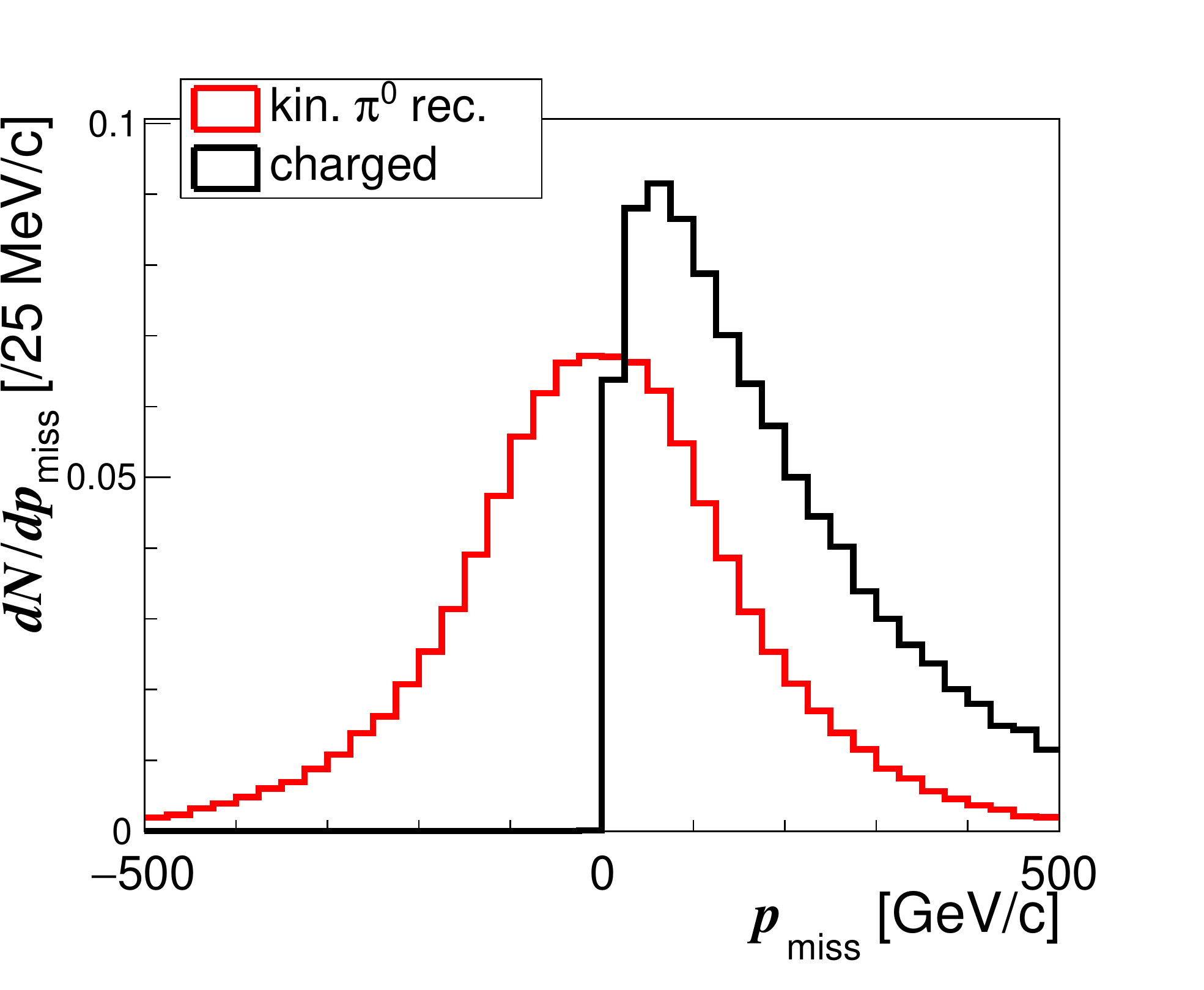}} &
\subfigure[]{\includegraphics[width=0.65\columnwidth]{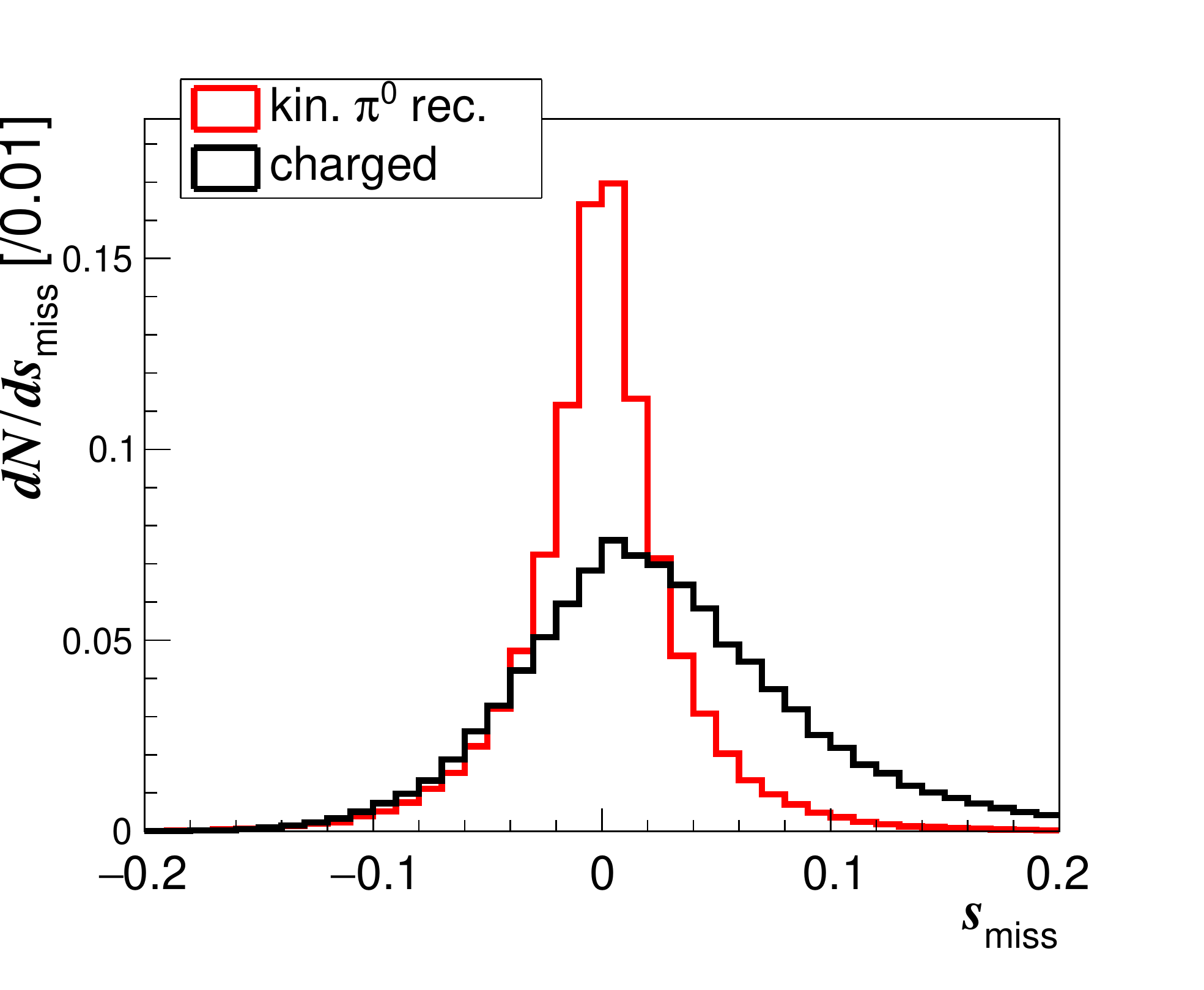}} &
\subfigure[]{\includegraphics[width=0.65\columnwidth]{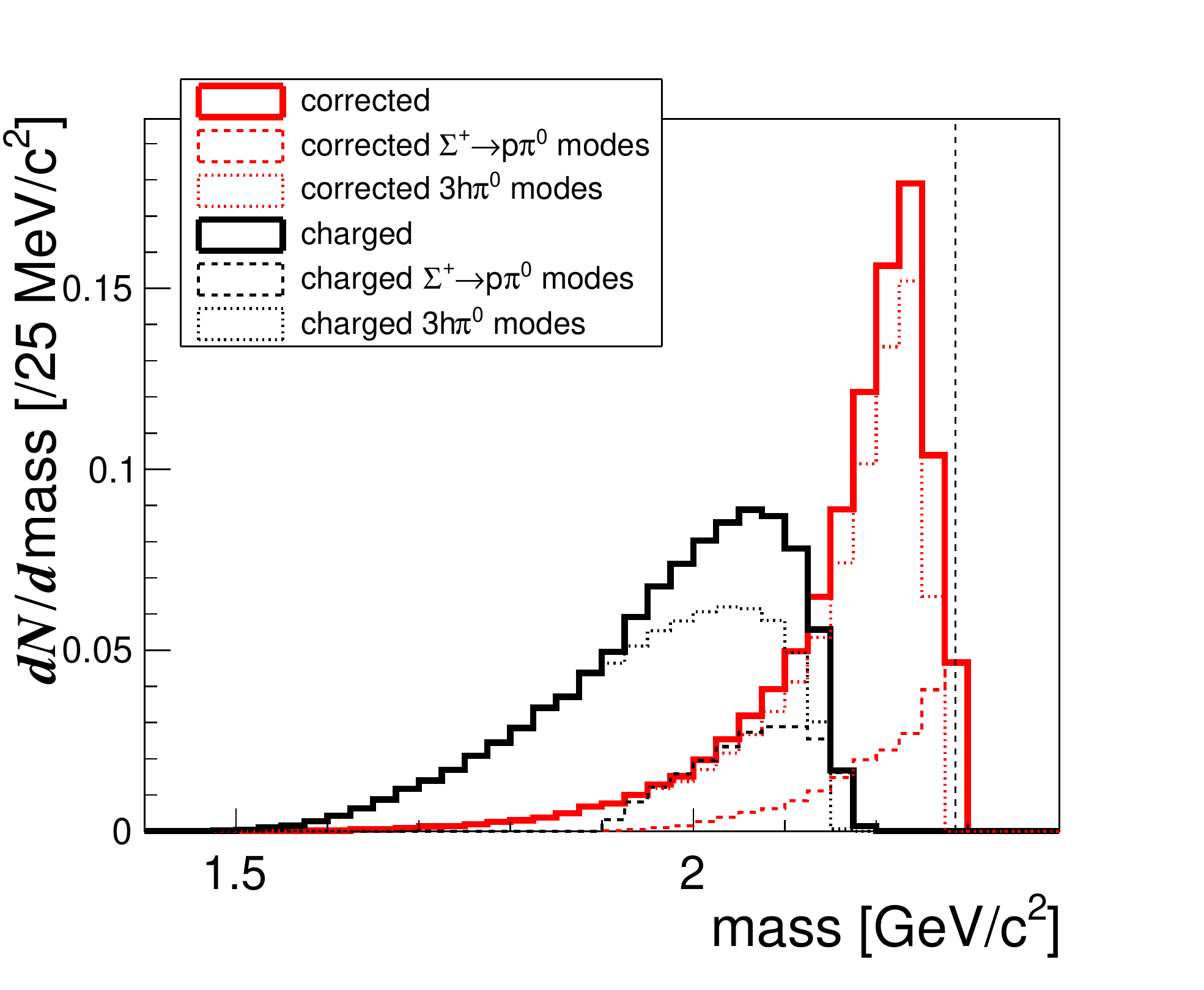}}
\end{tabular}}
\caption{\label{fig:neutrals}   
  Distributions of missing (a) momentum and (b) spin-polarization magnitudes using the kinematical reconstruction technique
for \Lc decays with an undetected neutral pion (red) in the final states listed in Table~\ref{tab:Lc_decay_modes}.
As a comparison, the missing momentum and polarization magnitudes for the reconstructed $3h$ system (black), composed by a proton combined with charged kaons and pions, are shown.
The relative contributions from the different final states are given by the effective branching fractions ${\cal B}_{\textrm{eff}}$.
(c) Corrected mass distribution for the same decays. As a comparison the invariant mass of the reconstructed charged $3h$ system, composed by a proton or a $\Sigma^{\pm}$ hyperon
combined with charged kaons and pions, is shown.
The \Lc known mass is reported as a vertical dashed line. The estimated invariant mass resolution at \lhcb, about 15~\mevcc, is negligible in comparison to the width of the corrected mass distribution.
All distributions are normalized to unity.
}
\end{center}
\end{figure*}

Figure~\ref{fig:neutrals}(a)(b) illustrates the difference between the estimated momentum and spin-polarization magnitudes using the kinematical recovery technique
and the true \Lc momentum, compared to the case when the \Lc three-momentum and polarization vectors are estimated using the $3h$ system only.
Whereas the former provides unbiased estimates with resolutions about $150\gevcc$ and $0.05$, respectively, the latter is clearly biased.
Estimated vertex resolutions around $100\mum$ and $10\mm$ in the transverse and
longitudinal directions, respectively, along with an invariant mass resolution of about $15\mevcc$~\cite{Bagli:2017foe}, do not impact significantly these distributions.

To characterize efficiently and with low background signal \Lc decays with undetected neutral pions, the corrected mass~\cite{Abe:1997sb}, defined as  
\begin{equation}
\label{eq:corrected_mass}
m_{\textrm{corr}}=\sqrt{m^2_{3h}+p^2_\perp}+p_\perp,
\end{equation}
is used, where $m_{3h}$ is the reconstructed invariant mass of the three charged tracks in the final state and $p_\perp$ is
the momentum of the $3h$ system transverse to the charm baryon flight direction, estimated as
\begin{equation}
\label{eq:3momentumreco}
p_\perp = | \boldsymbol p_{3h} \times \boldsymbol u | .
\end{equation}
Figure~\ref{fig:neutrals}(c) shows the $m_{\textrm{corr}}$ distribution for the mixtusre of three- and four-body \Lc decay modes listed in
Table~\ref{tab:Lc_decay_modes} with a \piz in the final state, and compared with the invariant
mass of the $3h$ system and the known \Lc baryon mass~\cite{Tanabashi:2018oca}. 
\subsection{Sensitivity studies and crystal setup optimization}
\label{sec:methods_sensitivity}
Sensitivity studies for the MDM and EDM of \Lc and \Xicp baryons
are based on \pythia simulations and 
pseudoexperiments
where the number of events is estimated
according to the branching fractions reported in Table~\ref{tab:Lc_decay_modes}
and Table~\ref{tab:Xic_decay_modes}, respectively. 
The values for the production cross sections,
the average event information $S^2$,
and the gyromagnetic factors are reported in
Table~\ref{tab:sensitivity}.
Other relevant parameters used for the sensitivity studies, also reported,
are evaluated for each different setup configuration according to simulations,
in particular: the channeling efficiency $\epsilon_\textrm{CH}$, the decay flight efficiency $\epsilon_\textrm{DF}$,
the number of recontructed charm baryons $N_{\textrm{rec}}$,
the average boost $\langle \gamma \rangle$ and transverse momentum $\langle \pt \rangle$,
and the initial average spin-polarization vector ${\boldsymbol s}$.
For the pseudoexperiments, Eq.~(\ref{eq:eom_sxsy_full}) is used for spin-polarization projections after precession in the crystal.

Regions of minimal uncertainty of $d$ and $g$ factors
are explored for different crystal configurations and target thickness, with
deflection angles of 16 and $7\mrad$.
These steering angles are chosen for compatibility with the \lhcb detector acceptance and to allow effective separation of channeled \Lc baryons from forwardly
produced background particles in the \Sb scenario~\cite{Bagli:2017foe}, respectively.
Crystal lengths around 10 and $7\cm$ are chosen to minimize material for detector operation and safety while accepting a 50\% and below 20\% increase of uncertainties, respectively.
The crystal orientation angle and target thickness are chosen to be $0.3\mrad$ and $2\cm$, respectively.

\begin{figure}[htb!]
\begin{center}
\subfigure[]{\includegraphics[width=0.75\columnwidth]{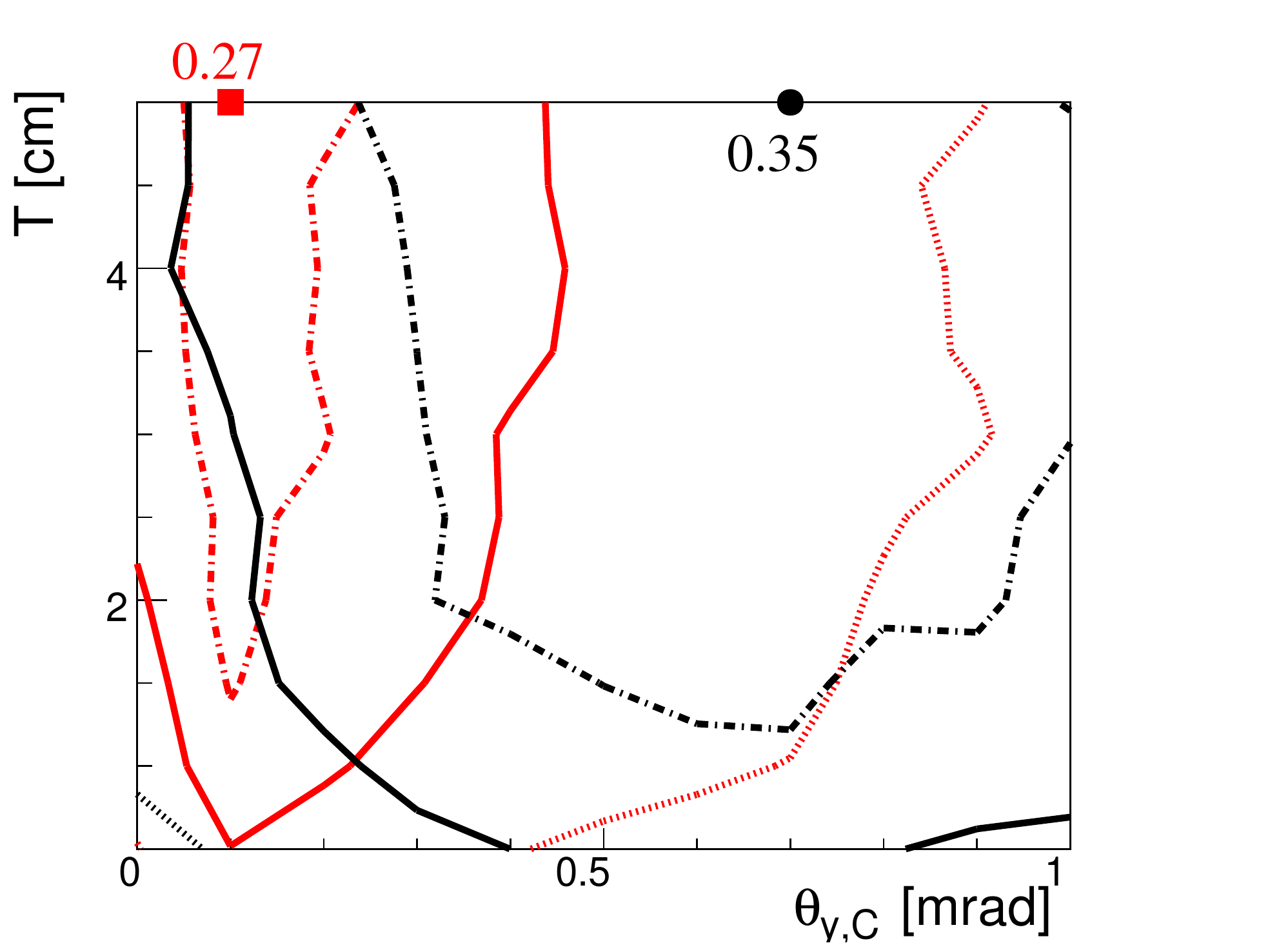}} \\  
\subfigure[]{\includegraphics[width=0.75\columnwidth]{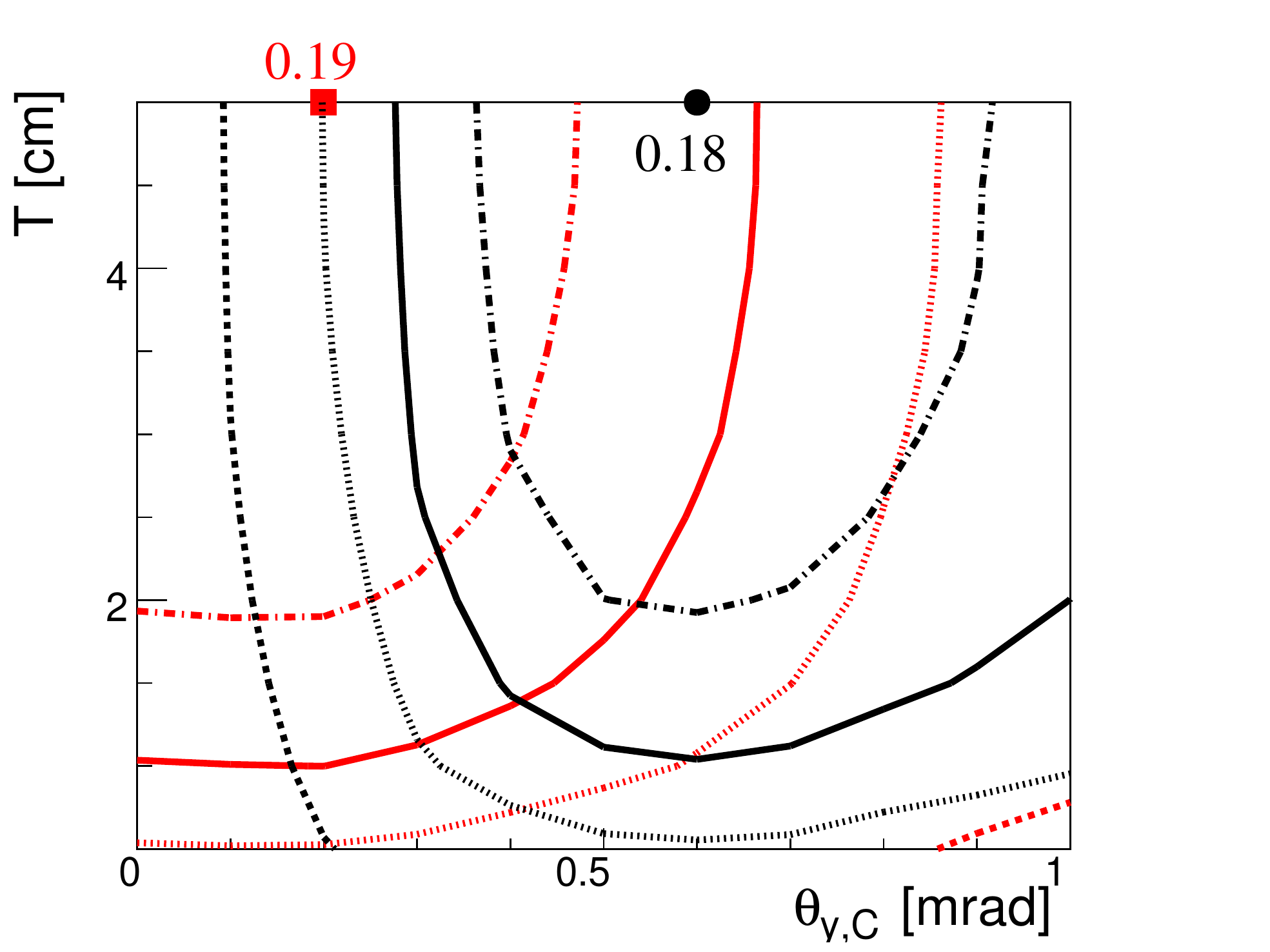}}
\caption{Regions of miniminal uncertainty of the $g$ (red curves) and $d$ (black) factors as a function of the crystal orientation angle $\theta_{y,C}$ and the target
thickness $T$, with $a \approx-0.03$ and $\Lc\to\pr\Km\pip$ decays,
(a) for combined MDM and EDM measurement using average estimates of the precesion angle $\varPhi$ and the initial polarization-vector ${\boldsymbol s}$, and 
(b) from first order, analytical estimates as given by Eq.~(\ref{eq:analytical_sensi}).
The markers and values represent the minimum
uncertainty on the $g$ and $d$ factors, relative to $1.37\times 10^{13}$ \pot
and a 16\mrad bent, 10\cm long germanium crystal at room temperature.
The curves are the regions
whose uncertainties are increased by 20\%, 50\% and 100\% with respect to the minimum.
\label{fig:optimization_thyCvsT_averageinfo}
}
\end{center}
\end{figure}

To assess the benefits of using event-by-event information, 
Fig.~\ref{fig:optimization_thyCvsT_averageinfo}(a) shows sensitivity regions as a function of the crystal orientation
angle and the target thickness for combined MDM and EDM measurement, using average event information.
These regions and minimal uncertainties are to be compared to those illustrated in Fig.~\ref{fig:optimization_thyCvsT}(b).
A sensitivity improvement equivalent to a data sample size larger than a factor of three (of two for MDM measurement alone) is
observed when using event-by-event estimates of the precesion angle $\varPhi$ and the initial polarization-vector ${\boldsymbol s}$
with respect to the case when these are averaged over all events.

First order, analytical approximation to the dipole moment sensitivities is given by
\begin{align}
\sigma_g & \approx \frac{2}{S \langle s_y \rangle \langle \gamma \rangle \theta_C} \frac{1}{\sqrt{N_{\rm rec}}}, \nonumber \\  
\sigma_{d} & \approx \frac{g-2}{S} \frac{1}{\sqrt{\langle s_y \rangle^2 \eta^2 + 2 \langle s_x \rangle^2 \eta}}
\frac{1}{\sqrt{N_{\rm rec}}}, 
\label{eq:analytical_sensi}
\end{align}
with $\eta = 1 - \cos \langle \varPhi \rangle$,
which follows from Eq.~(\ref{eq:eom_sxsy_full}) in the limit $a \gg d,\ 1/\gamma$,
\begin{align}
s'_x & \approx   s_y \frac{d}{g-2} (1-\cos\varPhi) + s_x , \nonumber \\  
s'_y & \approx   s_y \cos\varPhi                   + s_x \frac{d}{g-2} (1-\cos\varPhi), \nonumber \\  
s'_z & \approx  -s_y \sin\varPhi                   + s_x \frac{d}{g-2} \sin\varPhi,
\label{eq:eom_sxsy}
\end{align}
where $\varPhi$ is given by Eq.~(\ref{eq:varphi}). 
In this case the uncertainties are estimated separately for events with positive and negative $p_{x_L}$, and then combined.
Figure~\ref{fig:optimization_thyCvsT_averageinfo}(b) shows the corresponding sensitivity regions. Differences with respect to Fig.~\ref{fig:optimization_thyCvsT_averageinfo}(a) are related to
the approximations in Eq.~(\ref{eq:eom_sxsy}) and the correlations between $g$ and $d$ in the combined measurement, neglected in the analytical estimates. The dependence of the EDM sensitivity with the crystal orientation angle is more evident in Fig.~\ref{fig:optimization_thyCvsT_averageinfo}(b) due to an incomplete and approximated use of the spin precession information.
%
%
%
\indent
\section*{Acknowledgements} 
We express our gratitude to our colleagues of the \lhcb collaboration, in particular to %
M.~Ferro-Luzzi, 
M.~Palutan,
C.~Parkes,
P.~Robbe
N.~Tuning,
and from CERN, in particular A.~Fomin, D.~Mirarchi, S.~Redaelli for 
stimulating discussions and very useful feedback. We would like to thank
S.~Forte and J.~Rojo for useful discussions on parton distribution functions.
A.~Sytov thanks
V.~Haurylavets and A.~Leukovich for providing help with \geant
simulations. 
We acknowledge support from INFN (Italy), MICINN and GVA (Spain), the ERC Consolidator Grants SELDOM G.A. 771642 and CRYSBEAM G.A. 615089. In addition, we acknowledge the CINECA
award under the ISCRA initiative for the availability of high performance
computing resources and support.



 \end{document}